\documentclass{jaa}
\usepackage{natbib}

\usepackage{graphicx}
\usepackage{aasmacros}
\usepackage{color}

\newcommand{\rsun}{$R_{\odot}$}  

\usepackage{graphicx}

\graphicspath{{./}{figures/}}

\begin{document}\sloppy

\title{Preparing for Solar and Heliospheric Science with the SKAO: An Indian Perspective}


\author{Divya Oberoi\textsuperscript{1,*}, 
Susanta Kumar Bisoi\textsuperscript{2},
K. Sasikumar Raja\textsuperscript{3}, 
Devojyoti Kansabanik\textsuperscript{1},
Atul Mohan\textsuperscript{4,5}, 
Surajit Mondal\textsuperscript{6} and
Rohit Sharma\textsuperscript{7}}
\affilOne{\textsuperscript{1} National Centre for Radio Astrophysics, Tata Institute of Fundamental Research, S.P. Pune University Campus, Ganeshkhind, Pune 411007, India.\\}
\affilTwo{\textsuperscript{2} Department of Physics and Astronomy, National Institute of Technology, Rourkela 769008, India.\\}
\affilThree{\textsuperscript{3} Indian Institute of Astrophysics, II Block, Koramangala, Bangalore-560 034, India.\\}
\affilFour{\textsuperscript{4} Rosseland Centre for Solar Physics, University of Oslo, Postboks 1029 Blindern, N-0315 Oslo, Norway.\\}
\affilFive{\textsuperscript{5} Institute of Theoretical Astrophysics, University of Oslo, Postboks 1029 Blindern, N-0315 Oslo, Norway.\\}
\affilSix{\textsuperscript{6} Center for Solar-Terrestrial Research, New Jersey Institute of Technology, 323 M L King Jr Boulevard, Newark, NJ 07102-1982, USA.\\}
\affilSeven{\textsuperscript{7} Fachhochschule Nordwestschweiz,
Bahnhofstrasse 6, 5210 Windisch, Switzerland.\\}


\twocolumn[{

\maketitle

\corres{div@ncra.tifr.res.in.in}

\msinfo{1 January 2015}{1 January 2015}

\begin{abstract}
The Square Kilometre Array Observatory (SKAO) is perhaps the most ambitious radio telescope envisaged yet. It will enable unprecedented studies of the Sun, the corona and the heliosphere and help to answer many of the outstanding questions in these areas. Its ability to make a vast previously unexplored phase space accessible, also promises a large discovery potential. The Indian solar and heliospheric physics community have been preparing for this science opportunity.
A significant part of this effort has been towards playing a leading role in pursuing science with SKAO precursor instruments.
This article briefly summarises the current status of the various aspects of work done as a part of this enterprise and our future goals.
\end{abstract}


}]


\doinum{12.3456/s78910-011-012-3}
\artcitid{\#\#\#\#}
\volnum{000}
\year{0000}
\pgrange{1--}
\setcounter{page}{1}
\lp{1}

\section{Introduction} \label{sec:intro}
As in practically all other areas of astrophysics, the Square Kilometer Array Observatory (SKAO) is envisaged to be a powerful tool to dramatically enhance our understanding in the field of solar and heliospheric physics. Both within India and internationally, the scientific community pursuing this area of research at radio wavelengths, has historically been comparatively small, and that continues to be the case even today. Nonetheless, there has been a significant international effort to pursue this science opportunity with the SKAO which Indian researchers have also been a part of \citep{Nakariakov2015, Nindos2019}. In fact, the SKAO Science Working Group for Solar, Heliospheric, and Ionospheric (SHI) was set up at the Indian initiative. 

Indian solar and heliospheric science community is particularly well poised for playing a leading role in this area.
Being home to the Ooty Radio Telescope \citep[ORT;][]{SWARUP1971} and the set of instruments at the Gauribidanur Radio Observatory \citep[GRO;][]{Ram1998, Ram2011} which have been carrying out dedicated observations in this field and the Giant Metrewave Radio Telescope \citep[GMRT;][]{Swarup_1991,Gupta_2017} which has also been used for solar studies \citep[e.g.][]{Mercier2006,Mercier2015,Bisoi_2018}, India has had a rich and long tradition of solar and heliospheric studies.
At metrewave bands, the Murchison Widefield Array \citep[MWA;][]{lonsdale2009, Tingay2013,  Wayth2018} is currently perhaps the best-suited radio interferometer for providing high dynamic range and high-fidelity spectroscopic snapshot imaging required for meeting the science needs of solar and heliospheric science.
Ever since its inception, Indian scientists have been playing a leading role in establishing solar and heliospheric science as one of the key science objectives of the MWA and continue to lead it \citep{Oberoi2004, Oberoi2010, Bowman2013, Beardsley2020}.
Though the bulk of the work by the Indian community has been at frequencies overlapping with the SKAO-Low, the future plans involve making use of the SKAO-Mid frequencies as well.

The MWA is a precursor for the SKAO, especially the SKAO-Low.
Not only is it located at the site chosen for the SKAO-Low the MWA also shares many design features and hence also the challenges with the SKAO-Low.
The MWA was also among the first aperture array interferometers to be built and marks a big step in both engineering and science.
The data from the MWA are sufficiently different in their character and data volumes from most other existing instruments and their analysis represents a significant challenge in its own right.
Building reliable and precise tools and techniques for analyzing the MWA data lie directly on the path to doing the same for the SKAO.
As will be substantiated later, the confluence of these data and our specialized tools are already enabling explorations of previously inaccessible phase space.
These science explorations serve many different purposes, from academic to societal.
From a purely academic perspective, they deliver exciting science and help us sharpen our science questions for the SKAO.
They also provide us with very useful experience in preparing for solar and heliospheric science in the SKAO era even as we make good progress with the available instrumentation.
From an Indian perspective, they play a crucial role in developing the academic human resource needed in the country, not only to be able to pursue science with the SKAO once it becomes available but hopefully play a defining role in this science area.
By demonstrating cutting-edge science, these works also serve the purpose of attracting young talent to the field of solar and heliospheric radio physics.

This work gives an overview of our journey thus far, it summarises the current status and our plans for the near future. The paper is organised broadly into sections on solar and heliospheric physics, followed by a concluding section.

\section{Solar Physics} \label{sec:solar-phy}
The Sun is a particularly challenging radio source to be imaged. 
The challenges stem from the wide range of variations seen in solar emissions -- time scales spanning solar cycles to sub-second; spectral scales from smooth thermal emission to $\sim$100 kHz; brightness temperatures from  $10^4$K for gyrosynchrotron (GS) emission from coronal mass ejection (CME) plasma to $10^{13}$K for bright type-III solar bursts;  fractional polarisation ranging from $\leq$1\%  to nearly 100\%; and angular scales small enough to be unresolved by most earlier solar interferometers to the degree scale quiet solar disc and even larger CMEs \citep[e.g.][etc.]{McLeanBook,Sastry_2009,Nindos2020}. 
This made it infeasible for most radio interferometers to gather the information needed to make high dynamic range high-fidelity images over the short time and narrow spectral scales necessary to track the evolution of solar emission in detail.
Fortunately, however, for the active emissions from the well-known solar type-I through type V bursts, which can outshine the quiescent Sun by orders of magnitude, much of the physics can be learned from the characteristics of the emission seen in their dynamic spectra (DS) and imaging was not strictly necessary \citep{McLeanBook,Saint-Hilaire2013,Reid2014}. 
Hence much of what we have learned thus far about the various solar type bursts comes from a large body of literature based on non-imaging DS-based studies.
The majority of these studies have been limited to Stokes I measurements, though a small fraction has also included measurements of Stokes V emissions.
Certain information can, however, only be obtained using imaging studies -- estimates of brightness temperatures ($T_B$); disentangling the multiple sources of emission simultaneously present on the Sun; gathering morphological information of the emission features and their relationships with features observed in other wavebands, etc., 
Ideally, the imaging needs to be done with spectral and temporal resolutions finer than the frequency and time scales of the intrinsic solar variability, provide the sufficient dynamic range to simultaneously observe sources with vastly different intrinsic $T_B$s and also provide full Stokes information with leakages much smaller than the lowest expected fractional polarization.

While science-rich, interferometric imaging is very compute-intensive and spectro-polarimetric snapshot imaging pushes the demands on computational requirements by multiple orders of magnitude.
At meter wavelengths, there are additional challenges to be overcome for solar imaging, especially when using aperture array instruments like the MWA (and the future SKAO-Low), which add to its algorithmic complexity and computational demands \citep{mondal2019,Kansabanik2022b, Kansabanik2022c, Kansabanik2022d}.
These factors severely limit the volumes of data that can be studied in detail using spectroscopic snapshot imaging.
The use of DS is, on the other hand, much more straightforward.
Its computationally light and data volumes friendly nature allows it to be easily scaled to large data volumes.
Both imaging and non-imaging studies, hence, have their importance and complementary roles to play in solar physics. 
Enabling imaging studies with the new generation instruments necessarily requires the availability of a robust and capable polarimetric imaging pipeline.

The earliest work from the 32-tile prototype for the MWA already reported evidence for numerous weak and short-lived nonthermal emissions, even during comparatively quiet solar conditions \citep{Oberoi2011}.
Working towards characterizing their observational features and building an understanding of these physical systems has been an enduring theme of our work.
This section describes the major aspects of our work towards non-imaging studies, the development of interferometric imaging pipelines tailored to the needs of solar radio imaging, and the imaging studies enabled by them.

\subsection{Non-imaging Studies}  \label{sec:non-imaging-studies}
As discussed earlier, non-imaging studies are technically much simpler as compared to imaging studies, and form the natural place to start when pursuing science with any new instrument.
We also took the same approach with the MWA.

\subsubsection{Flux Density Calibration} \label{DS-flux-density-calibration} 
Reliable flux density calibration has always been a non-trivial issue for solar radio observations, especially at low radio frequencies. 
For dedicated solar instruments, there are only a handful of sources in the sky, which have sufficient flux density to be 
used as flux density calibrators. Additional complications arise due to the confluence of the bright Galactic background and the typical large fields of view (FoV) of low-frequency instruments.
For instruments designed for usual astronomical observations, the vast difference in the flux densities of the Sun and the typical calibrator sources usually pose a challenge for the linearity range of the electronic signal chain and need special attention. 

For the MWA, we developed a novel approach to provide accurate solar flux density calibration.
The use of the Galactic background for calibrating an array of active elements had already been demonstrated \citep{Rogers2004}, and we extended this work to the case of a two-element interferometer \citep{Oberoi2017}.
The necessary ingredients for this technique are -- a reliable model for the Galactic background over the MWA band; an accurate model for the MWA primary beams; a robust characterization of the MWA receiver; and baselines lengths long enough to resolve out the bulk of the Galactic emission, but short enough to view the Sun as an unresolved source. These were all available from prior work and/ or the MWA design. 

This computationally lean technique uses $<0.1$\% of the MWA data and can provide absolute flux density calibration with an accuracy of 10$-$60\%. The accuracy of relative flux density calibration is $\sim1$\%. This technique is equally applicable to quiet and active times and can provide calibration at the native time and frequency resolution of the data. It played a key role in establishing flux densities of many weak solar radio bursts \citep[e.g.][etc.]{Suresh2017,Sharma2018} and a prescription was successfully developed and implemented to translate this flux density to MWA radio images \citep{Mohan2017}.
This combination has since been used successfully in many of the studies described in Secs. \ref{sec:bursts} and \ref{sec:CME-gyro}.

\subsubsection{Wavelet-based Identification and Characterization of Features in Dynamic-spectrum} \label{sec:wavelets} 
Early MWA observations led to an appreciation of the presence of numerous short-lived and narrow-band emission features even during periods of comparatively low solar activity. The strengths of these features were much weaker than those typically associated with the well-known solar bursts \citep{Oberoi2011}.
The natural next step was to characterize these emissions in terms of the various parameters which could be discerned using the DS from MWA observations.
Their high observed rate, of thousands of emission features per hour in the 30.72 MHz of MWA bandwidth, necessarily required an automated but robust approach for accurately identifying and parameterizing these features.
\citet{Suresh2017} developed a wavelet-based automated technique to reliably identify and parameterize these emission features under weak to moderate solar activity conditions.
Though the implementation presented was tuned to MWA data, the technique itself is general and can be used on DS from other instruments, including the future SKAO-Low.

Some pre-processing of the raw MWA DS was needed to prepare the data for wavelet-based feature detection.
After removing the known instrumental artifacts, a mean background continuum level was estimated and subtracted from each of the 4-minute chunks of data.
A Gaussian Mixtures Model (GMM) approach was used for background estimation.
GMM provides a probabilistic framework based on the assumption that the data come from a parent distribution consisting of a superposition of a finite numbers of Gaussian and uses the data to estimate the parameters of
these constituent Gaussians 
\citep{McLachlan_GMM,Pedregosa2012}.
The DS data were modeled as a collection of Gaussians and, because of the low to medium levels of solar activity prevailing during these observations, the Gaussian component with the lowest mean and largest weight was assumed to represent the thermal background continuum.
A suitable low-degree polynomial was fit to the smooth spectral trends observed and subtracted from the observed data leaving behind residuals in the 3--4\% range.
A 2D Ricker wavelet, also known as a Mexican hat, was used as the mother wavelet in the Continuous Wavelet Transform (CWT) framework to first identify the locations of these features and then also estimate their spectral and temporal extents.

\begin{figure}[!t]
    \includegraphics[trim={0cm 0.4cm 0cm 0.5cm},clip,scale=0.29]{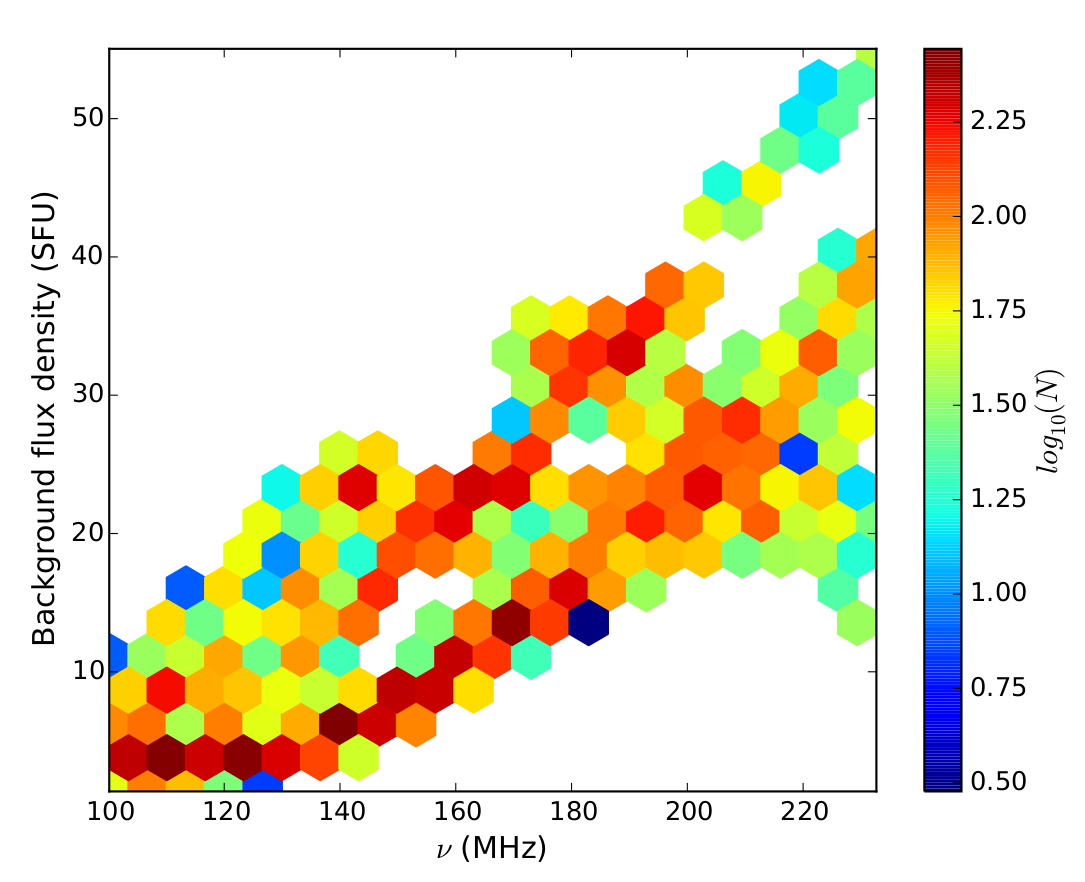}
    \centering
    \caption{{\bf{Variation of the background solar flux density.}} The 2D histogram highlights the variation in the background solar flux density seen at the frequency and time, where an emission feature was identified in DS using the wavelet based analysis. The colour axis is on a log scale. (Reproduced from \citet{Suresh2017})}
    \label{fig:wavelet-work}
\end{figure}

This technique detected about 14,200 impulsive emission features in about 4.5 hours of observations with flux densities in the range from 0.6--$\sim$300 SFU (1 SFU = 10$^4$ Jy), making these perhaps the weakest nonthermal emissions reported till then.
A power law fit over the 12--155 SFU range yielded a power law index, $\alpha=-2.2$ for peak flux densities.
Their energies were estimated to lie in the range $10^{15}$--$10^{18}$ ergs, much lower than typical ranges for solar type-I and type-III radio bursts, and described well by a power slope of $-1.98$.
The distributions of their bandwidths and durations were found to be smooth and unimodal with peaks around 4--5 MHz and 1--2 seconds respectively.
These emission features seemed to be fairly uniformly distributed across the spectral range probed by these observations.
While the spectral profiles were usually found to be symmetric about the peak, the temporal profiles showed no such symmetry.
Interestingly, the background flux densities at the location in DS of these features are found to vary by a factor of $\sim$2, as shown in Fig. \ref{fig:wavelet-work}.
These large variations are not likely to arise due to changes in the thermal emission from the coronal plasma.
The authors interpret this as evidence for the presence of multiple instances of nonthermal emission which remained unresolved in the observations.

\subsubsection{Gaussian Mixture based Characterization of Weak Nonthermal Impulsive Emissions}\label{sec:gauss_mix} 

\begin{figure}[!t]
\includegraphics[trim={1.5cm 0.1cm 1.5cm 0cm},clip,scale=0.48]{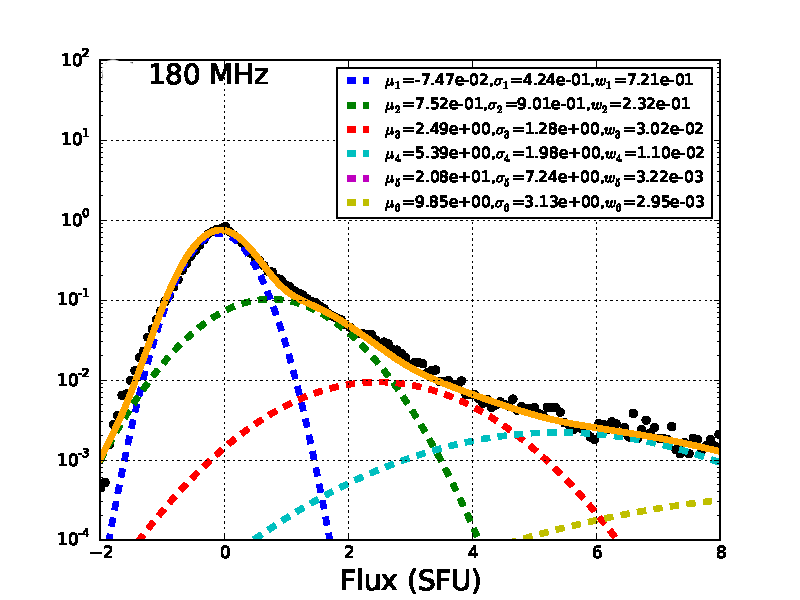}
\centering
\caption{{\bf{Gaussian mixture modeling of solar emission.}} The observed distribution of solar ﬂux density at 180 MHz for 70 minutes of observation is shown by black dots in units of SFU.
The component Gaussians are shown by dashed lines and their sum by the solid orange line.
Note that the y-axis is in log scale and spans four orders of magnitude.
For this dataset, the GMM found 7 Gaussians components, the means ($\mu$), widths ($\sigma$), and weights ($w$) are listed. (Reproduced from \citet{Sharma2018})
}
\label{fig:GMM-work}
\end{figure}

It is reasonable to regard the emission seen in the solar DS as a superposition of a slowly varying (order minutes and longer) and impulsive nonthermal emission (order seconds and smaller).
The slowly varying emission is expected to be dominated by the coronal thermal bremsstrahlung, while the impulsive emissions must arise from nonthermal processes in the corona.
The widespread presence of the nonthermal impulsive features was obvious even in the science commissioning data from the MWA (2013). 
Estimating their flux densities and coming up with metrics to quantify their presence, therefore, became one of the first natural questions to answer and was addressed by \citet{Sharma2018}.
A dataset with a medium level of activity was chosen for this exercise.
The flux density calibration of a DS from a short baseline was done as described in Sec. \ref{DS-flux-density-calibration}. 
To separate the impulsive emissions from the slowly varying component a robust median filter over a 120 s window was used and the median so computed was subtracted from the observed DS.
As the impulsive nonthermal emissions are superposed on the background thermal emission, they must always appear as a positive tail in the histogram of the resulting DS of impulsive emission.
An example histogram of impulsive emission is shown in Fig. \ref{fig:GMM-work} along with the various component Gaussians determined using GMM analysis.
The Gaussian with zero mean represents the slowly varying thermal emission and, for periods of medium solar activity, also has the largest weight amongst all Gaussian components.
The most interesting aspect of this work was that it provided a way for quantifying both the prevalence and the flux density of impulsive nonthermal emissions.
In, what was a surprising result at the time, we found that the prevalence (fractional occupancy) of the nonthermal emissions ranged between 17 and 45 \% and that the energy emitted in the impulsive emissions is very comparable to that emitted in the slowly varying component even during periods of medium levels of solar activity.
The density of impulsive emissions was also found to correlate with the strength of the slowly varying emission, similar to what is seen for Type-I bursts.
The authors suggested that the observed increase likely arises due to impulsive emissions too fine to be resolved by these observations.

\subsection{Development of Calibration and Imaging Pipelines}
Solar radio images are a rich source of information, going well beyond what DS can offer.
However, for reasons enumerated at the start of Sec. \ref{sec:solar-phy}, it is an intrinsically hard problem.
In addition, especially during periods of solar activity, very faint emissions co-exist (e.g. GS and thermal emissions from the CME plasma) with very bright active emissions (e.g. type-II, type-IV bursts, and noise storms).
To understand the overall coronal dynamics and magnetic field topology, these emissions, spanning a huge range of $\mathrm{T_B}$, need to be detected in the same image. This requires high imaging dynamic range high fidelity spectro-polarimetric snapshot imaging.

The unique array layout of the MWA, with a large number of antenna elements distributed over a small footprint, makes it exceptionally well suited for high dynamic range spectroscopic snapshot imaging of the Sun. Due to its large number of antenna elements, the MWA data is already very voluminous, about a TB of raw data every hour, and this number will increase many-fold with the availability of the SKAO.
The need for spectroscopic snapshot imaging translates to making a very large number (tens to hundreds of thousands) of images at high time and spectral resolutions, rendering the traditional manual data analysis approach impractical.
While several groups across the world are working towards automating interferometric imaging procedures in the run-up to the SKAO, the unique requirements of the solar observations make it hard to adapt the generic techniques developed primarily for astronomical observations for the solar use case. 
The Indian Solar Physics community has been putting in a concerted effort over the past few years toward building robust calibration and imaging pipelines optimized for low-frequency solar observations. 
Not only can these pipelines operate in an unsupervised manner, but the images from these pipelines also show improvements by multiple orders of magnitude in dynamic range over earlier attempts and achieve polarimeteric leakages on par with high quality imaging of other astronomical objects. They now represent the state-of-the-art in low radio frequency solar imaging.

While the current implementation of these pipelines has been optimized for the MWA, due to similarities in array architecture with the SKAO, especially the SKAO-Low, it will be possible to easily adapt these pipelines for solar imaging with the SKAO. 
Key design considerations for these pipelines were to, on the one hand, hide the bulk of the complexity by providing well-chosen defaults for the numerous parameters which need to be specified and thus make high-quality solar radio imaging accessible to the uninitiated, and on the other, provide all of the flexibility to the expert user to tune the calibration and imaging parameters as desired.
Successive incarnations of these pipelines are described in the following text and have met the stated objectives to a large extent.
Our ultimate objective is to provide the community of solar scientists with the tools which would enable them to generate 
science ready solar images at meter wavelengths like the current practice for visible, extreme ultra-violet (EUV), and X-ray band data from the Solar Dynamics Observatory \citep[SDO;][]{Pesnell2016}, Solar and Heliospheric Observatory \citep[SOHO;][]{Domingo1995} etc.

\subsubsection{Total Intensity Calibration and Imaging Pipeline}

The first incarnation of this pipeline was named  ``Automated Imaging Routine for Compact Arrays for Radio Sun'' \citep[AIRCARS;][]{mondal2019}. 
Most of the early MWA solar work was based on non-imaging studies (Sec. \ref{sec:non-imaging-studies}). 
While these studies were interesting in their own right, they were only scraping the surface in terms of making use of the information content of the MWA data, both in terms of using its imaging capability and sensitivity.
The fact that interferometric imaging is intrinsically an iterative process and the traditional analysis methods are very human effort intensive, made it infeasible to generate the very large number of images needed for spectroscopic snapshot studies.
Additionally, early efforts to adapt the existing generic astronomical imaging tools to solar imaging led to imaging dynamic ranges much lower than those expected from the MWA data.

A key reason behind this was the deficiencies in the calibration. 
At the MWA, calibrator observations were generally done when the Sun was below the horizon to avoid contamination from the solar emission. 
Although the antenna themselves have a very stable instrumental response \citep{Sokolwski2020}, the ionospheric phase usually changed significantly between the calibrator and target (Sun) observations and rendered the calibration solutions obtained from night-time observations sub-optimal for solar observations.
While a similar issue is regularly tackled in radio imaging using a technique called self-calibration, solar observations are generally not amenable to it due to the lack of good initial calibration solutions available for solar observation and the highly non-linear nature of the self-calibration process. 
AIRCARS overcomes this by using the large number of MWA antennas and its centrally condensed architecture cleverly by generating a sufficiently precise though low-resolution solar model for initializing the self-calibration process. 
For any given time slice, AIRCARS starts with calibration of only the phases of antenna gains using all baselines with at least one of their antennas from the dense central MWA core. 
Antennas outside the core are progressively added. 
The antennas from the compact MWA core look through essentially the same ionosphere and provide a certain level of coherency of the array \citep{Kansabanik2022c}. This allows AIRCARS to initiate the self-calibration process with a reasonable solar model.
AIRCARS also incorporates additional heuristics to tune the entire process so that the calibration solutions converge to their true values.

\begin{figure}[!t]
    \centering
    \includegraphics[trim={0cm 0.3cm 0cm 0cm},clip,scale=0.67]{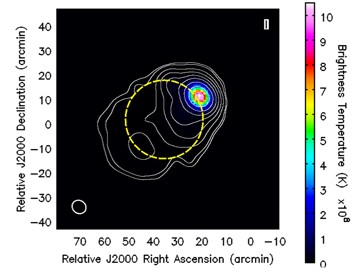}
    \caption{{\bf{A sample image from AIRCARS.}} Solar image during a type-II radio burst is shown. The image is made over 40 $\mathrm{kHz}$ and 0.5 $\mathrm{s}$. Peak $T_B$ is $\sim10^9\ \mathrm{K}$. Dynamic range of the image is 100,000. Contour levels are at 0.02, 0.03, 0.1, 0.3, 1, 3, 10, 30, 80, and 90 \% of the peak $T_B$. Dotted yellow circle represents the optical disc of the Sun (Reproduced from \citet{mondal2019})}.
    \label{fig:aircars_showcase}
    \vspace{-1cm}
\end{figure}

AIRCARS has been extensively used with data from a wide variety of solar and ionospheric conditions \citep{mohan2019a,mohan2019b,mondal2020a,mondal2020b,mondal2021a,mondal2021b,mohan2021a,mohan2021b,Kansabanik2022a} and has produced some of the highest dynamic range images at these frequencies. 
It routinely produces images with dynamic ranges $\gtrsim300$ for the quiet featureless Sun and $\gtrsim10^5$ in presence of bright and compact active solar emissions. 
A sample image demonstrating the imaging quality of AIRCARS is shown in Fig. \ref{fig:aircars_showcase}. The image shows a type-II solar radio burst in progress with peak $\mathrm{T_B}$ $\sim 10^9\ \mathrm{K}$. 
The thermal emission from the solar disc is at $\mathrm{T_B}\approx10^5\ \mathrm{K}$ and is detected with high significance. For this image, the imaging dynamic range is $\sim10^5$. 

\subsubsection{Polarimetric Calibration and Imaging Pipeline :}
\label{sec:P-AIRCARS}

Having established the ability of MWA solar observations to deliver high dynamic range high fidelity Stokes I images using AIRCARS, we have recently developed the second incarnation of AIRCARS. 
Referred to as ``Polarimetry using Automated Imaging Routine for Compact Arrays for Radio Sun'' \citep[P-AIRCARS,][]{Kansabanik2022b,Kansabanik2022c, Kansabanik2022d}, it is designed for robust polarisation calibration and imaging of the solar observations with the MWA. 

Full Stokes polarisation calibration is more challenging than working only with Stokes I. These challenges are even greater for the low-radio frequency full Stokes solar imaging. Solar emissions can span a very large range of intrinsic polarisations ($\leq1\%$ to $\sim100\%$) which can also vary rapidly across time and frequency. Additionally, the wide field-of-view (FoV) aperture arrays tend to have large direction-dependent instrumental polarization.
Due to these reasons, some of the assumptions made for routine polarimetric calibration for small FoV instruments are no longer valid in this regime \citep{lenc2017}.

Polarimetric solar observations with the MWA have been carried out successfully earlier, though only for the brighter active emissions, and using an ad-hoc approach of mitigating instrumental polarisation \citep{Patrick2019,Rahman2020}. 
The limited imaging dynamic range and polarimetric fidelity it can deliver make this method unsuitable for weak emissions and those with a low degree of fractional polarisation. 
P-AIRCARS overcomes these limitations by implementing a robust polarisation calibration. For low radio frequency solar observations with aperture arrays, it is not feasible to obtain calibrator observations at nearby times with the same primary beam pointing as used for target (Sun). P-AIRCARS is, hence, designed to work with any calibrator observation. It builds on three pillars -- i) self-calibration; ii) availability of a reliable instrumental beam model \citep{Sokowlski2017}; and iii) some well-established properties of low-frequency solar radio emission.
Conventional polarisation calibration algorithms available in standard radio interferometric packages (e.g. {\sc CASA} and {\sc AIPS}) assume the instrumental leakages to be small and are hence not suitable for polarisation self-calibration of the MWA solar observations. 
P-AIRCARS uses a full Jones matrix calibration formalism implemented in {\it CubiCal} \citep{cubical2018}.
In addition to polarisation calibration, P-AIRCARS also incorporates several improvements beyond AIRCARS.
These include bandpass self-calibration, a more flexible framework, and optimizations based on learnings from the AIRCARS experience. These have led to improvements in Stokes I imaging quality over AIRCARS. 
P-AIRCARS is the state-of-the-art polarisation calibration and imaging algorithm for low-frequency solar observations. 
Though it is currently optimized for MWA solar observations, P-AIRCARS will be equally suitable for the future SKAO, especially the SKAO-Low. 

\begin{figure}[!t]
    \centering
    \includegraphics[scale=0.33]{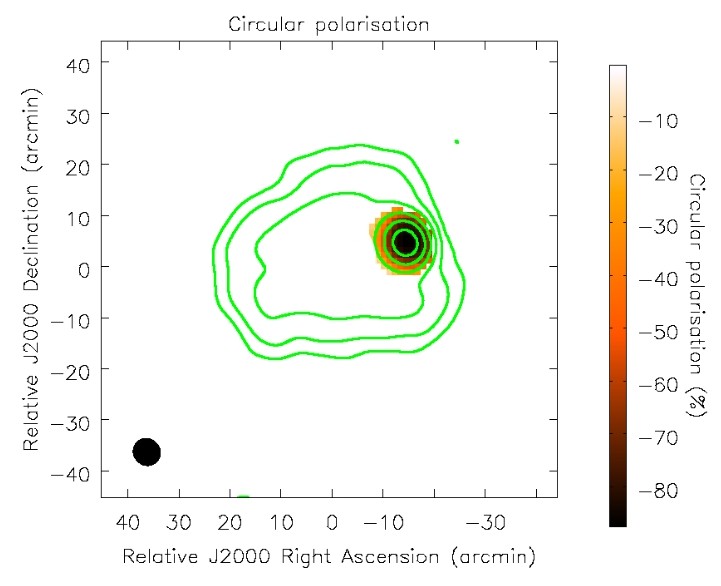}
    \caption{{\bf{A sample image from P-AIRCARS.}} A circular polarisation image of a type-I noise storm at 159 MHz is shown by the color map. The image is made over 40 $\mathrm{kHz}$ and 0.5 $\mathrm{s}$. The dynamic range of Stokes I and Stokes V images are $\sim600$ and $300$ respectively. The Peak degree of circular polarisation is approximately $-88\%$. Green contours show the Stokes I emission. Contour levels are at 5, 10, 20, 40, 60, and 80 \% of the peak of the Stokes I emission.}
    \label{fig:paircars_showcase}
    \vspace{-0.5cm}
\end{figure}

An example P-AIRCARS image is shown in Fig. \ref{fig:paircars_showcase}, which comes from a time when a type-I noise storm was in progress. 
P-AIRCARS routinely achieves residual instrumental leakages $\leq|2|\%$ from Stokes I to other Stokes parameters.
For this particular example, residual leakage from Stokes I to Stokes V is $<|1.8|\%$. These values of residual leakages are comparable to what is generally achieved for high-quality astronomical observations with the MWA \citep{lenc2017,lenc2018,Risley2018,Risley2020} and represent a significant improvement over what has been reported earlier for solar observations \citep{Patrick2019,Rahman2020}. 
The Stokes images delivered by P-AIRCARS are limited primarily by the thermal noise of the data.
The precise polarisation calibration provided by P-AIRCARS will make it possible to routinely detect and estimate the plasma parameters of the CMEs using the polarisation images of the gyrosynchrotron \citep[e.g.][etc,]{bastian2001,mondal2020a} or thermal emission \citep[e.g.][etc.]{Gopalswamy1993,Ramesh2021} from the CME plasma, and measure the very low level of circular polarisation ($\leq|1|\%$) \citep{Sastry_2009} from the quiet Sun thermal emissions, which remains a standing challenge. 
With the SKAO sensitivity, it should become feasible to detect the low level of circular polarisation and measure the global coronal magnetic field routinely, even during the quiescent periods.   

\subsubsection{Image-based Flux Density Calibration}
Absolute flux density calibration of the low-frequency solar observations has several challenges, as already discussed in Sec. \ref{DS-flux-density-calibration}. 
The same section also presented a non-imaging absolute flux density calibration technique developed for the MWA \citep{Oberoi2017}, which has been used in several studies.
Though successful, this technique relied crucially on the availability of multiple very short baselines ($\lesssim10\lambda$) to obtain the total solar flux density, which restricts the adaptability of this approach for the extended configuration of the MWA Phase-II \citep{Wayth2018}. AIRCARS and P-AIRCARS can provide high dynamic range images without requiring any dedicated calibrator observations. But due to the intrinsic limitation of the self-calibration based approach used in AIRCARS and P-AIRCARS, they cannot provide an absolute flux density calibration. 

With the precise calibration and high dynamic range imaging capability using AIRCARS and P-AIRCARS, it is possible to detect numerous background galactic and extra-galactic radio sources even in the presence of the Sun in the FoV (Fig.  \ref{fig:other_sources}). 
This presents the opportunity to arrive at an image-based absolute flux density calibration by comparing the observed flux densities of these sources with their catalog flux densities obtained from the GLEAM survey \citep{Wayth2015,walker2017}.
This approach is demonstrated convincingly by \citet{Kansabanik2022a}.
This work also used the well-established stability of the MWA bandpass shapes \citep{Sokolwski2020} to establish the robustness of the image-based absolute flux density calibration.
The key achievement of this work is that it provides the framework to arrive at absolute flux density calibration for any MWA solar observation, irrespective of the array configuration used.
It delivers an uncertainty of $\sim$10\% on the estimated flux density, which arises mostly due to the intrinsic uncertainty of the GLEAM flux density measurements. This method also demonstrates the usefulness of a well-characterized and the stable instrumental response for precise flux density calibration, a useful lesson for the design of the SKAO-Low.

The state-of-the-art full Stokes calibration and imaging pipeline, P-AIRCARS, and the robust absolute flux density calibration open a new window of high fidelity spectro-polarimetric snapshot imaging studies of the meter wavelength Sun. 
The success of these techniques and tools can be measured by the unprecedented quality of imaging they deliver.
They have been developed with the future needs of the SKAO in mind and can potentially form the backbone of the standard full Stokes solar imaging with the SKAO.



\subsection{Studies of Solar Radio Bursts} \label{sec:bursts}
Solar radio bursts are among the brightest observable radio transients - lasting from a few seconds to several hours, with $T_B$ ranging from $\sim 10^7$ -- $10^{13}\ \mathrm{K}$ \citep{McLeanBook, Saint-Hilaire2013}. 
They have been studied extensively using both imaging and non-imaging techniques over several decades \citep[see][for a review]{Reid2014}. Based on the emission morphology in the frequency-time plane (DS), solar radio bursts are classified into several types starting from types I through V \citep{Wild1950}. 
Though the early classification has stood the test of time for some seventy years, every significant advancement in instrumentation capabilities has brought about discoveries of previously unappreciated aspects of emissions at finer scales, often leading to elaborate sub-classification of each burst type.
The most recent wave of such discoveries is in progress today owing to instruments like the MWA and LOFAR.
\citep[e.g.][]{Suresh2017,Kontar2017,sharykin2018_LOFAR_dnn_withtypIIIb,mohan2019a,mohan2019b,Magdalenic2020,mohan2021b,mondal2021b}. 
Of the various solar burst types, we focus on the ones which had garnered a lot of attention in the recent past primarily owing to some notable discoveries.

\subsubsection{Type-III Solar Radio Bursts}
Type-III bursts are caused by supra-thermal electron beams produced at particle acceleration sites in the corona. These beams traverse across open magnetic field lines threaded across coronal iso-density layers. Along their path, they excite the two-stream instability which leads to a series of wave-wave and wave-particle interactions that give rise to coherent plasma emission from the iso-density layers at the local plasma frequency ($\nu_p$) and its harmonic frequency($2\nu_p$) \citep[e.g.][]{ginzburg1958,melrose1970,robinson_cairns1994}. Since $\nu_p$ relates to local density, by analyzing the mean type-III source locations in spectroscopic images, the radial density profile of the solar corona can be inferred \citep[e.g.][etc.]{Morosan2014,patrick2018_densmodel_frmtypIII}.

Type-III bursts are known to demonstrate fine-scale spectro-temporal variability in the image plane and intensity. Though an individual burst lasts for only a few seconds in a metrewave band, often groups of type-III bursts lasting for minutes to $\sim$ an hour are observed. The spectro-temporal intensity maps of these bursts demonstrate a variety of characteristics like quasi-periodic pulsations (QPPs), scattered striae like emission confined in frequency and time, skewed temporal profiles, etc. The rise and decay timescales of burst pulses follow power-law trends with observation frequency, with spectral indices close to -1 \cite[e.g.][]{Reid2014,Kontar19_Arznercopy}. 
Besides these features in the DS, the type-III burst sources also show interesting morphological characteristics in the image plane. These include variations in source sizes and centroid locations across frequency and time.
Some of these are related to the radio-wave propagation effects, while others are linked to the properties of particle acceleration processes in the corona. To disentangle these two effects, high-fidelity imaging at sub-second sub-MHz resolutions is necessary.

Limited imaging capabilities forced earlier studies exploring sub-second, sub-MHz spectro-temporal evolution of these bursts to rely purely on DS data. 
These studies had no choice but to tacitly attribute the observed variability in flux density to a single source of emission.
While this would usually be true for strongest events, this need not be the case for relatively weaker bursts with $T_B \sim 10^8 $--$10^{10} \mathrm{K}$.
In fact, weak type-IIIs in this $T_B$ range are well known to be associated with micro- or mini- flares  \citep[e.g.][]{benz2005_xray_beyC5_radioCor,Reid2017_typeIII-GOES-WIND_correlation}, which are in turn known to be very common especially during active periods \citep[e.g.][]{ash2012_flarestats}.
This suggests that the temporal overlap of multiple independent weak flares is a likely occurrence and limits the usefulness of DS studies for weaker events.

\begin{figure*}[!t]
    \centering
    \includegraphics[trim={0.1cm 0.1cm 0cm 0cm},clip,scale=0.6]{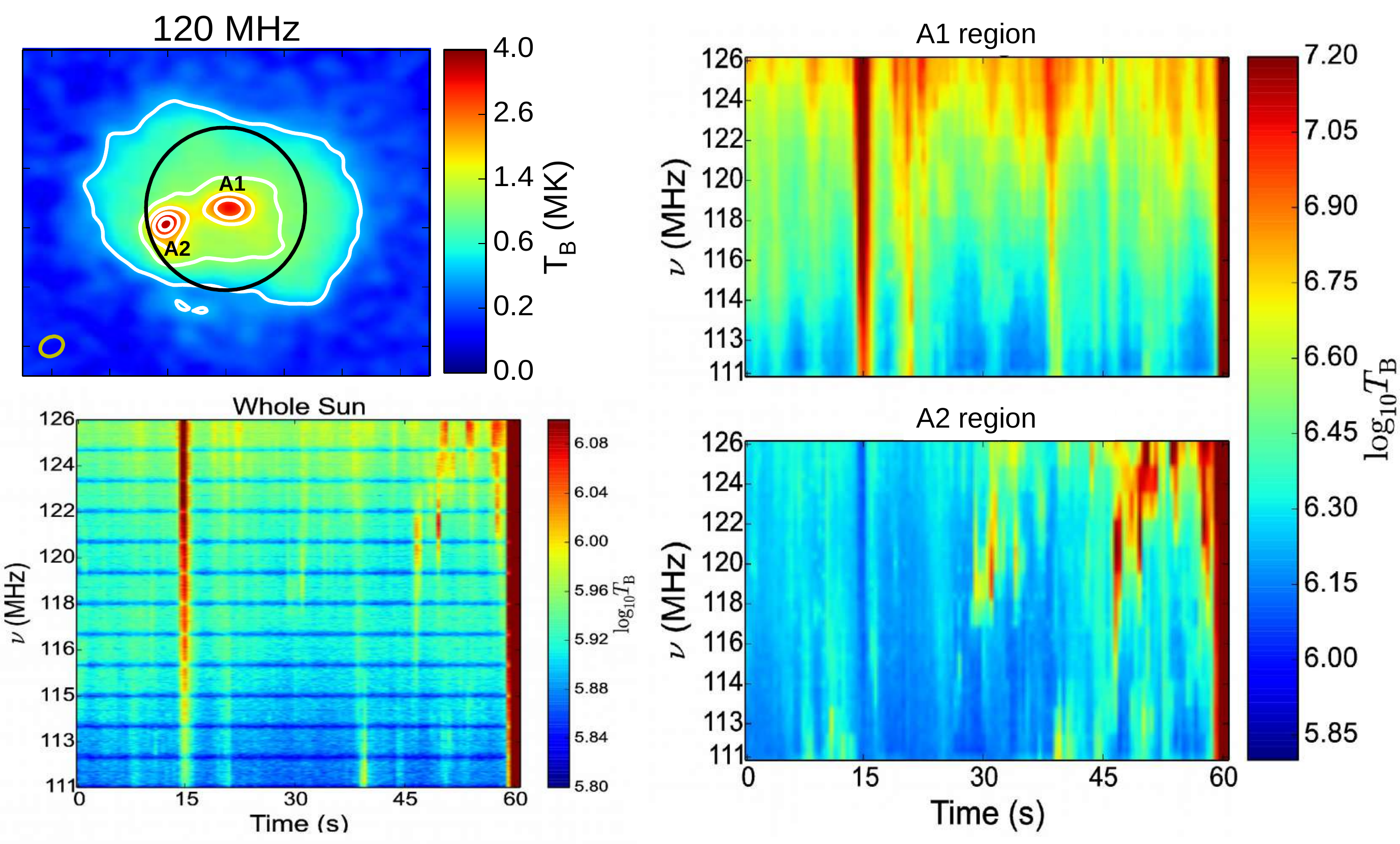}
    \caption{{\it Left panel: }Sample image from MWA with two active regions A1 and A2 (top) and the whole sun integrated dynamic spectrum for a minute-long observation (bottom). The black circle in the top image shows the optical disk of the Sun and the ellipse to the bottom left the resolution of the observation. Contours mark 10, 30, 50, 70 and 90\% of the peak $T_B$. {\it Right panel: }SPREDS for A1 and A2 revealing the localized emission features. (Adapted from \citet{Mohan2017})}
    \label{fig:SPREDS}
    \vspace{-0.4cm}
\end{figure*}

There have been several imaging studies of type-IIIs as well from earlier generations of instruments, though most of them also focused on strong type-III events which out-shined any co-temporal events by several orders of magnitude. 
These images were often made with integration times of few to many seconds and yielded dynamic ranges of only a few tens 
\citep[e.g.][]{kundu83_typeIII-ARTB,Saint-Hilaire2013}. 
The burst source regions undergo flux enhancements over several orders of magnitude during their short burst span ($T_B\sim 10^7 - 10^{12}\ \mathrm{K}$) as revealed by dynamic spectral studies \citep[e.g.][]{Saint-Hilaire2013}.

It is hence evident that to study weaker bursts which tend to overlap in time, and show rapid variability spanning multiple orders of magnitude in $T_B$ over fine spectro-temporal scales one necessarily needs high fidelity snapshot spectroscopic imaging with dynamic ranges of order 10$^3$. The new generation interferometric arrays like the MWA now make it possible.
The set of images spanning a range in time (observing duration) and frequency (bandwidth) with fine spectro-temporal resolution would generate a 4D data cube with information spread across frequency, time, and angular sky coordinates. 
To provide a convenient framework for studying such a datacube, in analogy with the usual DS, we defined SPatially REsolved Dynamic Spectra (SPREDS), which shows the time-frequency structure of emission arising from a given resolution element on the Sun. 
Figure \ref{fig:SPREDS} shows a sample image of the Sun along with SPREDS  for the two active regions A1 and A2. The whole sun dynamic spectra are also shown. 
The differences between the emissions from A1 and A2 are self-evident and there is no way to discern this from the usual DS.

\begin{figure}[!t]
    \centering
    \includegraphics[trim={0cm 0.3cm 23cm 0cm},clip,scale=0.29]{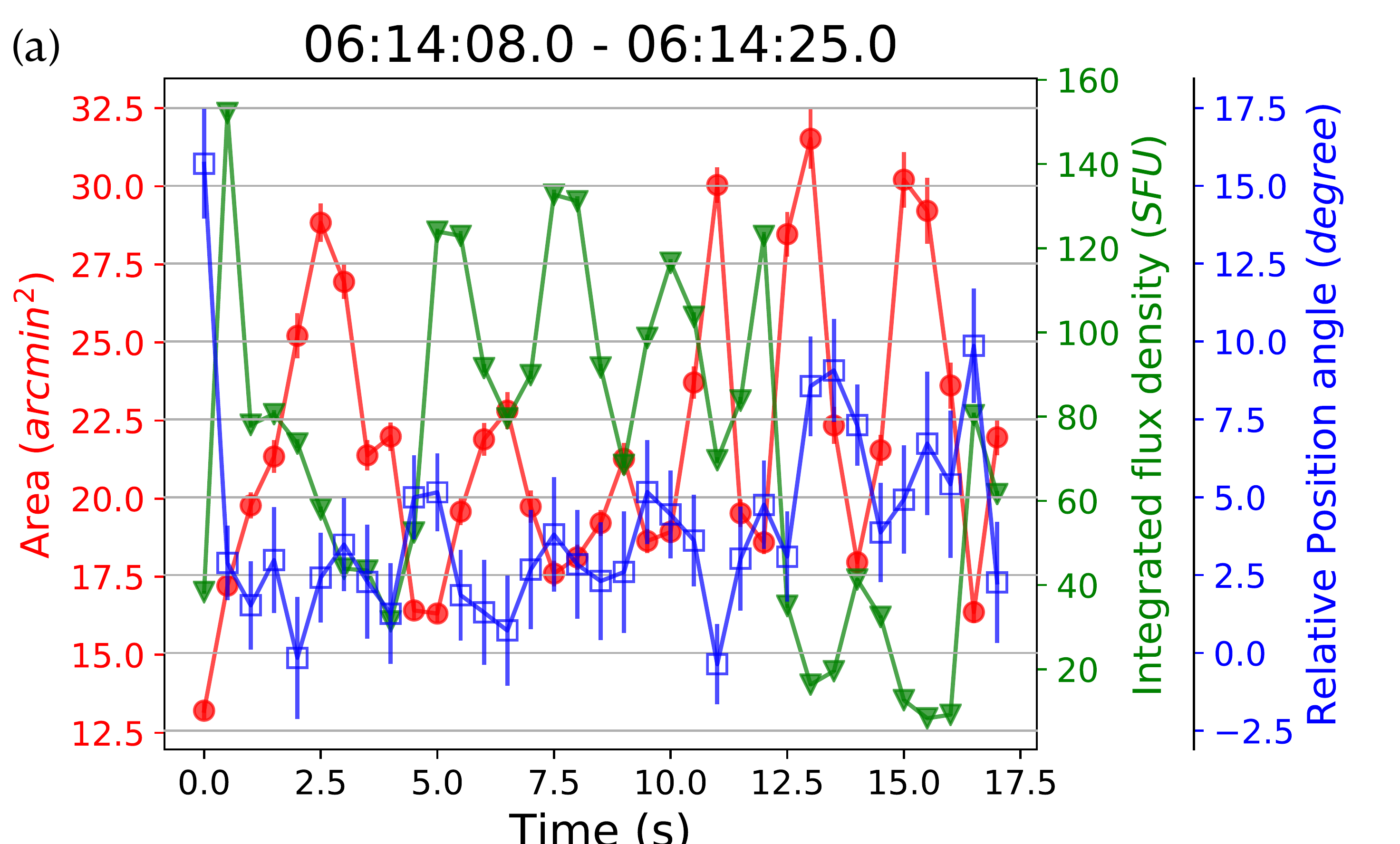}
    \caption{QPPs discovered in source area, flux density and orientation. The anti-correlation in time series of area of the type-III source and its integrated flux density, as derived from the best fit Gaussian parameters, is visually evident.
    (Adapted from \citet{mohan2019a})
    }
    \label{fig:typIIIEvol}
    \vspace{-0.1cm}
\end{figure}
\cite{mohan2019a} presented the first high dynamic range snapshot spectroscopic ($\delta t \sim 0.5\,s$; $\delta \nu \sim 160\,kHz$) imaging study of a group of weak type-IIIs using MWA images generated with AIRCARS. This group of bursts lasted only for a few minutes and was associated with a GOES B6 class flare.
A type-I source was also present, the $T_B$ of which was occasionally comparable to the type-III bursts.
The morphology of the type-III burst emission was always found to be described well by a 2D Gaussian function. 
The best fit 2D Gaussian parameters to type-III emission provided a detailed characterization of the spectro-temporal variability of the intensity and morphology of the type-III source during the event.
The availability of such fine-grained characterization enabled the authors to disentangle the intrinsic dynamics of the source from propagation effects through the inhomogeneous coronal medium.
This and other aspects related to propagation effects are discussed in Section \ref{sec:propeffect_bursts}. 

This study conclusively demonstrated the existence of a new type of QPP where the source sizes showed an anti-correlated pulsation with intensity, akin to a sausage mode at timescales of $\sim$ 2\,s. They also discovered similar timescale QPPs in source orientation. 
Figure \ref{fig:typIIIEvol}(a) shows these QPPs for one of the episodes from this group of bursts. 
While QPPs in the intensity of type-III sources were already known and three degenerate classes of models had been proposed to explain this phenomena \citep{Asch_QPPtheoryRev1987}.
Establishing the presence of QPPs along the area and orientation axes had to wait for high-fidelity imaging to become available.
The most important aspect of this work was that the additional information available from imaging enabled the authors to break the degeneracy in favor of the lone model where QPPs arise as a result of dynamics close to the particle acceleration site, possibly due to coupling with a local magnetohydrodynamic (MHD) wave mode \citep{ash94_pulsdAccl}.
The other two relied on phenomena relating to emission mechanism along the path of the electron beam and were ruled out as they required physical motion of magnetic flux tubes at speeds about two orders of magnitude larger than the estimated local Alfv\'{e}n speeds.

We have also used imaging studies of type-III events to derive 
the scales of coronal flux tubes across a heliocentric height range of 1.4 - 1.8\,$R_\odot$ using MWA data in 80 -- 200\, MHz range \citep{mohan2019a, mohan2021b}. The sizes derived were consistent with the empirical model proposed by \cite{Asch_observersview2003}. 
Type-III emissions are circularly polarised, there have been few imaging studies exploring this. 
The first such attempt has been made using the MWA data as well \citep{Rahman2020}, though it relied on a heuristic approach for polarimetric calibration \citep{Patrick2019}.
This work was already able to demonstrate previously unappreciated aspects, like significant polarisation fluctuations happening across different stages of a single event from its onset to decay.

These results demonstrate the scientific potential of snapshot spectroscopic imaging and the new phase space of research it has opened up. 
With the availability of our high fidelity polarimetric imaging pipeline (Sec. \ref{sec:P-AIRCARS}), a whole new axis of this phase space will become accessible for exploration.
We intend to make good use of this opportunity by continuing to identify and study interesting events and work towards building a general physical picture.

\subsubsection{Type-I Solar Radio Bursts}
Type-I bursts or the solar noise storms are another form of coherent plasma emission sources associated with active regions. Unlike the type III bursts, these events are not associated with any obvious particle acceleration events or flares. They can be associated with non-flaring active regions undergoing magnetic field reconfiguration \citep{smith62_flare_typeI_correlStats, iwai12_typeI_smallscaleEUVmagDyn_link}.

\begin{figure*}[!t]
    \centering
    \includegraphics[width=\textwidth, height=0.25\textheight]{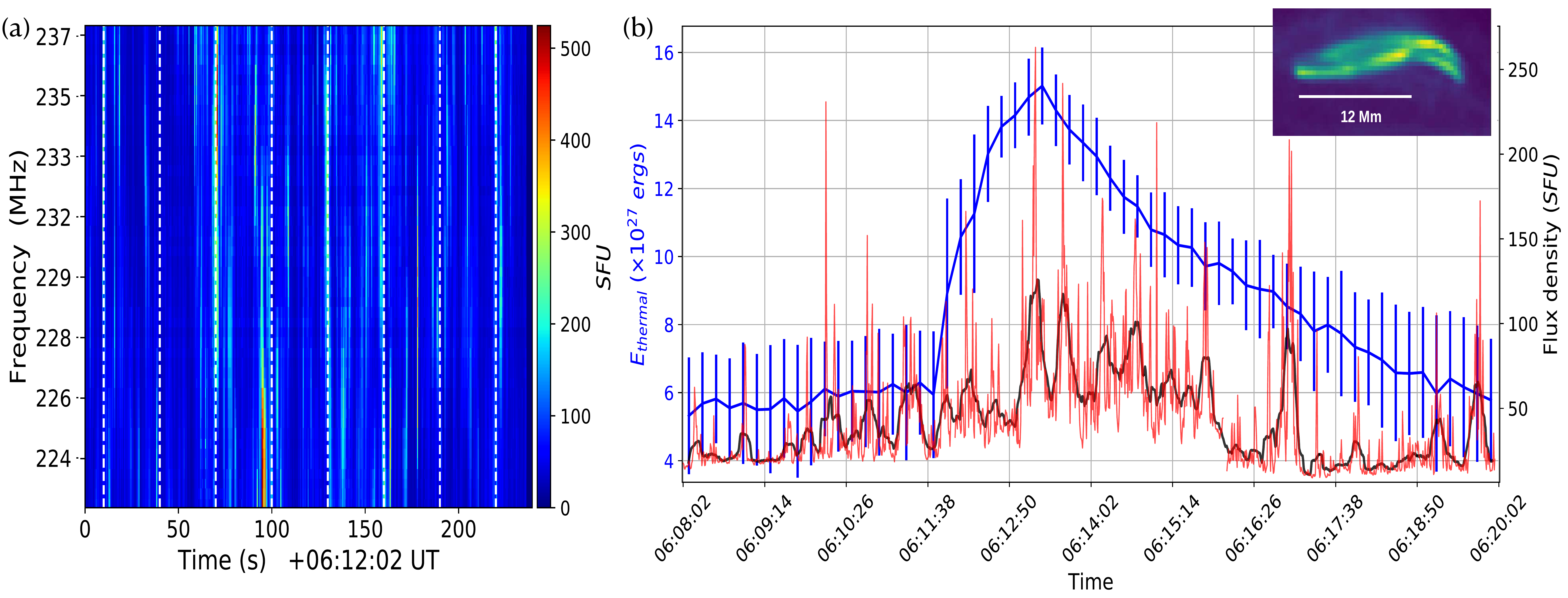}
    \caption{{\it (a): }SPREDS for the type-I source during the microflare. White lines marked at 30\,s intervals reveal the periodic bunching of the burst-like emission features. {\it (b): }Frequency averaged SPREDS data (red). The black curve shows the same data processed by a 10\,s running window to reveal the 30\,s QPPs. Thermal energy evolution derived from EUV data is shown in blue. The inset shows a EUV 94\AA\ image of the braided ARTB loop region.(Adapted from \citet{mohan2019b})}
    \label{fig:typIEvol}
    \vspace{-0.3cm}
\end{figure*}

Type-I bursts, unlike type IIIs, can last for hours and appear intermittently above active regions during different phases of their existence. They also show QPPs occasionally, the origins of which and the existence of their high energy counterparts are not well understood yet. 
These QPPs have longer periods, ranging from several seconds to minutes, unlike type IIIs, hinting at a different physical origin. Using SPREDS based studies \cite{mohan2019b} studied the spectro-temporal characteristics of a weak type I storm associated with an Active Region Transient Brightening (ARTB) event. 

\begin{figure*}
    \centering
    \includegraphics[trim={0cm 0cm 0cm 0cm},clip,scale=0.25]{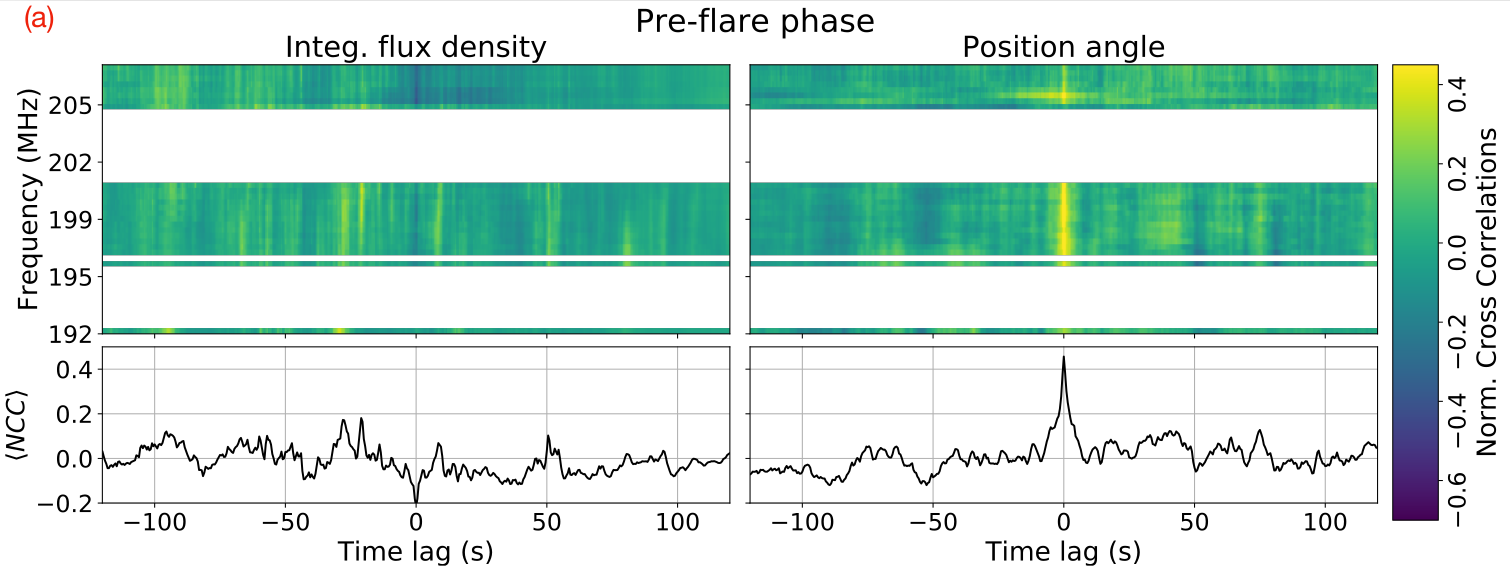}
     \includegraphics[trim={0cm 0cm 0cm 0cm},clip,scale=0.25]{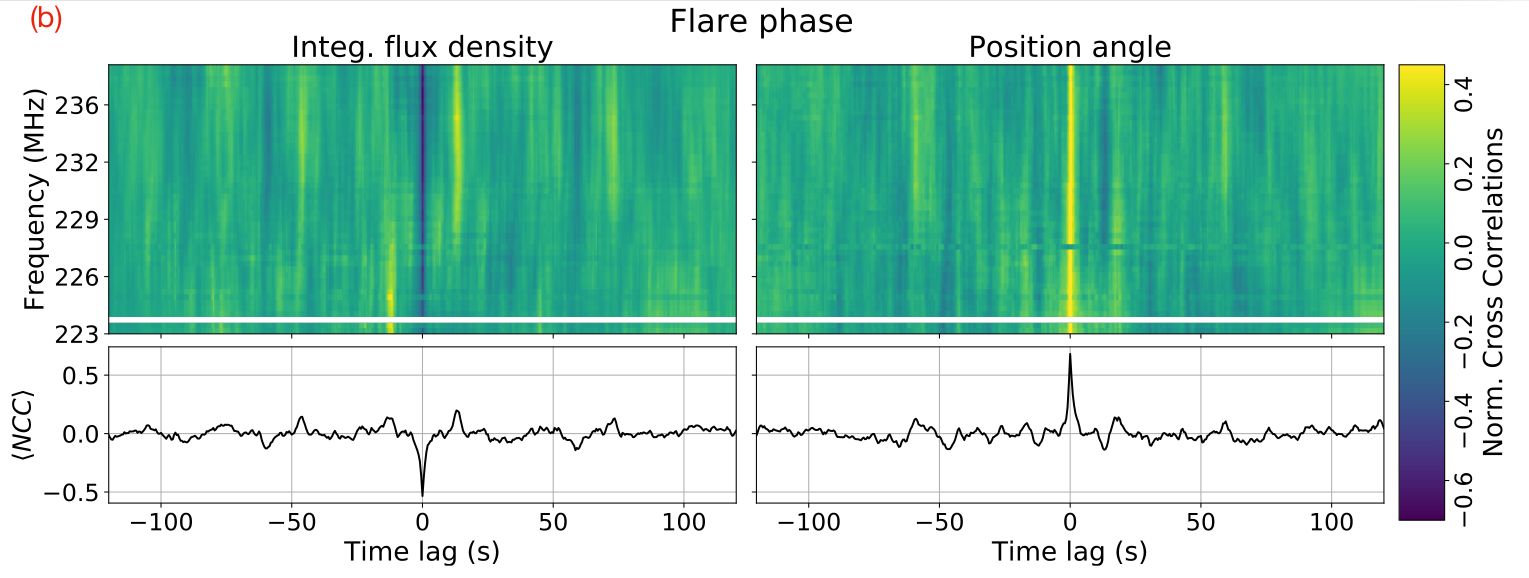}
      \includegraphics[trim={0cm 0cm 0cm 0cm},clip,scale=0.25]{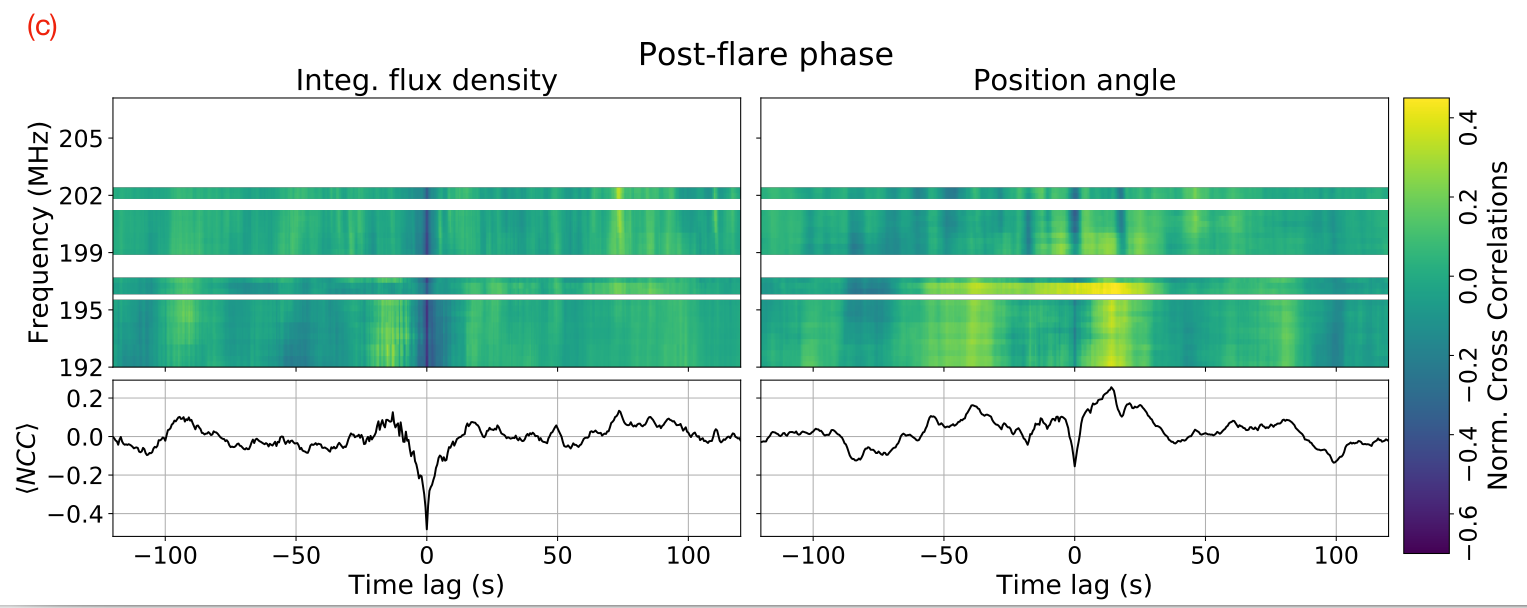}
       \includegraphics[trim={0cm 0cm 0cm 0cm},clip,scale=0.25]{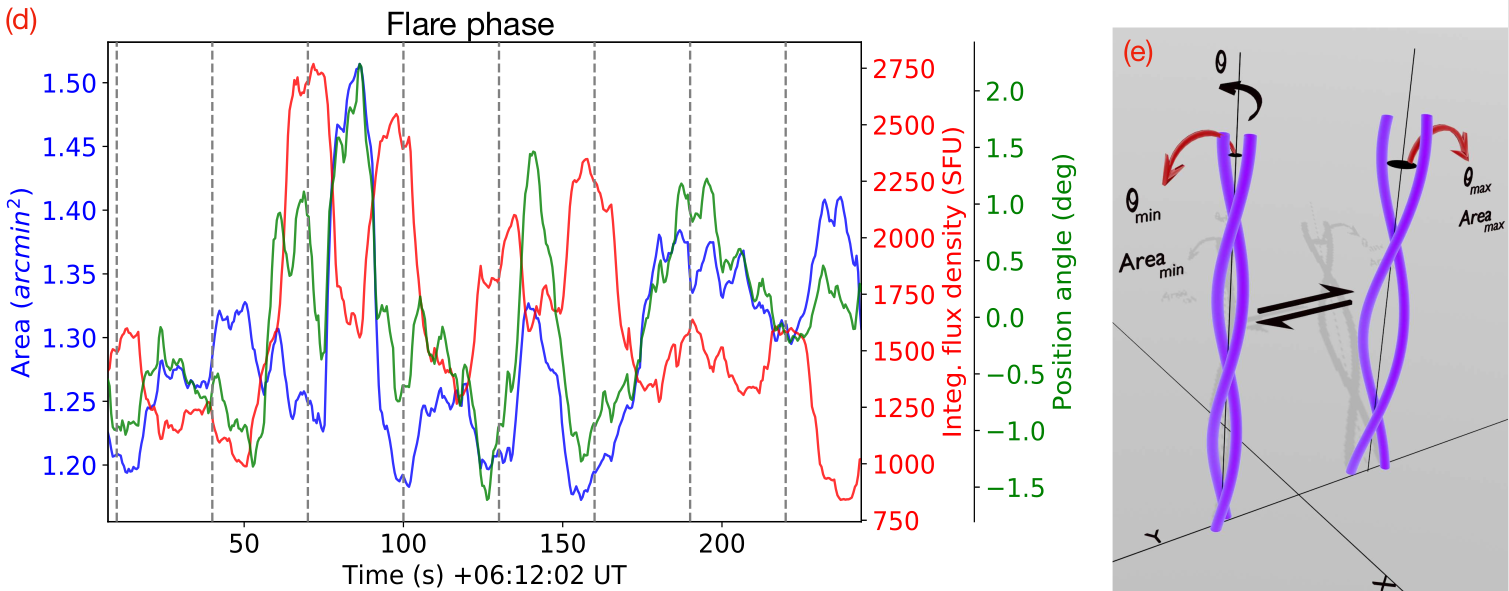}
    \caption{{\it (a-c): }Normalized cross-correlation functions for source flux versus area (left) and position angle versus area (right) for every observation frequency. Dominant correlated variation mode shifts from `T' to `S' during the event. {\it (d):} Correlated 30\,s QPPs in the band averaged SPREDS data which is processed by a 30\,s running mean filter. Vertical lines are marked every 30\,s. {\it (e): }Schematic of the `T' mode. (Adapted from \citet{mohan2021a} and \citet{mohan2021b})}
    \label{fig:typIQPP}
    \vspace{-0.3cm}
\end{figure*}
The ARTB was associated with a weak microflare. The study incorporated EUV and X-ray imaging analysis of the associated active region, which provided the thermal energy evolution. Magnetic field extrapolation studies using the Non-Linear Force Free technique \citep{wheatland07_nlfff} were carried out to understand the free energy content in the local magnetic field, which powered the microflare. This was the first detailed multiwavelength study of a type I--ARTB-- microflare system, from pre-flare to post-flare phase. The radio source was present in all three phases.
It revealed many interesting aspects including the presence of 30\,s QPPs in the radio data. QPPs were present before, during, and after the microflare, but the intensity and the regularity of the pulsations were enhanced during the flare (Fig.~\ref{fig:typIEvol}(b)) with $T_B$ approaching $10^9\,K$. 
The analysis showed that the QPPs resulted from a systematic release of magnetic free energy across the braided coronal loops at local Alfv\'{e}n timescales in the radio source heights. The lower coronal EUV source images revealed the presence of braiding in the region at the scales required to explain the QPPs (Fig.~\ref{fig:typIEvol}(b)). The magnetic free energy released during the QPPs produced numerous accelerated particle beams, signatures of which could be found in the SPREDS as type-III like strands of emission (Fig.\ref{fig:typIEvol}(a)). These beams were subsequently damped by collisions in the ambient medium.  

In an independent study, the structural evolution of the type I source was studied for the first time by \cite{mohan2019b}. The sources were found to be well described by 2D Gaussians.
This work led to the discovery of correlated modes of variability, often seen as pulsations, in the integrated flux density, area, and position angle of the source across all observational bands centered around 200\,MHz.
The correlated evolutionary modes can be classified into a sausage-like `S' mode in which the area evolves in an anti-phased manner with integrated flux density;  and a wind-unwind-like `T' mode in which the area evolves in a correlated manner with position angle (schematic - Fig. ~\ref{fig:typIQPP}(e)). Figure~\ref{fig:typIQPP}(d) shows the correlated QPPs in the flare phase.
In the pre-flare data, the presence of `T' mode was found but not `S', whereas it was the contrary in the post-flare phase. The flare phase marked the rise of the `S' mode alongside the existing `T' mode. This transition is seen in the normalized cross-correlation functions evaluated for the integrated flux density and position angle with respect to the area during the event in pre- to post-flare phases (Fig.~\ref{fig:typIQPP} (a-c)).
The `T' mode could be related to the presence of excess free energy in torsional modes in the excessively strained pre-flare magnetic loops driven by footpoint motions. Such a scenario has been long predicted by coronal loop simulations and this excess energy is expected to be released in the form of heating and in reorganizing of the local magnetic field structure. This can explain the observed microflare and the rise of `S' mode during the flare, releasing the excess `T' mode energy. As a result mode conversion happens via the flare from `T' to `S' mode (Fig.~\ref{fig:typIQPP} a-c).

Though the magnetic field evolution of internally twisted loops has been addressed by simulations \citep[e.g.][]{Gordov12_MagRelax_loops, Threlfall18_interacting_twisted_loops}, these could not be studied in the case of weak events. This is because the observations were primarily focused on the high energy and microwave bands, which are sensitive to only relatively strong flares which cause major changes in the local magnetic field structure \citep[e.g.][]{Gordovs12_Effect_XrayLC_internalReconn,Gordovs_20_XclassFlare_particleAcc}. This implies that the meterwave type-I source structural evolution studies, like the one presented, are a unique and powerful technique to track the dynamics of internally twisted loops with minimal thermal and nonthermal energy flux using the evolution of the coherent flux density from the high energy electron beams produced in the region. More such studies need to be done for a variety of active regions varying in magnetic field strengths, configurations, and other physical characteristics. This will help improve the understanding of the prevalence and characteristics of correlated QPPs, and their links to physical dynamics.

\subsection{Gyrosynchroton Emission from CMEs} \label{sec:CME-gyro}

\begin{figure*}[!t]
    \centering
    \includegraphics[trim={3cm 1.2cm 4cm 1cm},clip,scale=0.7]{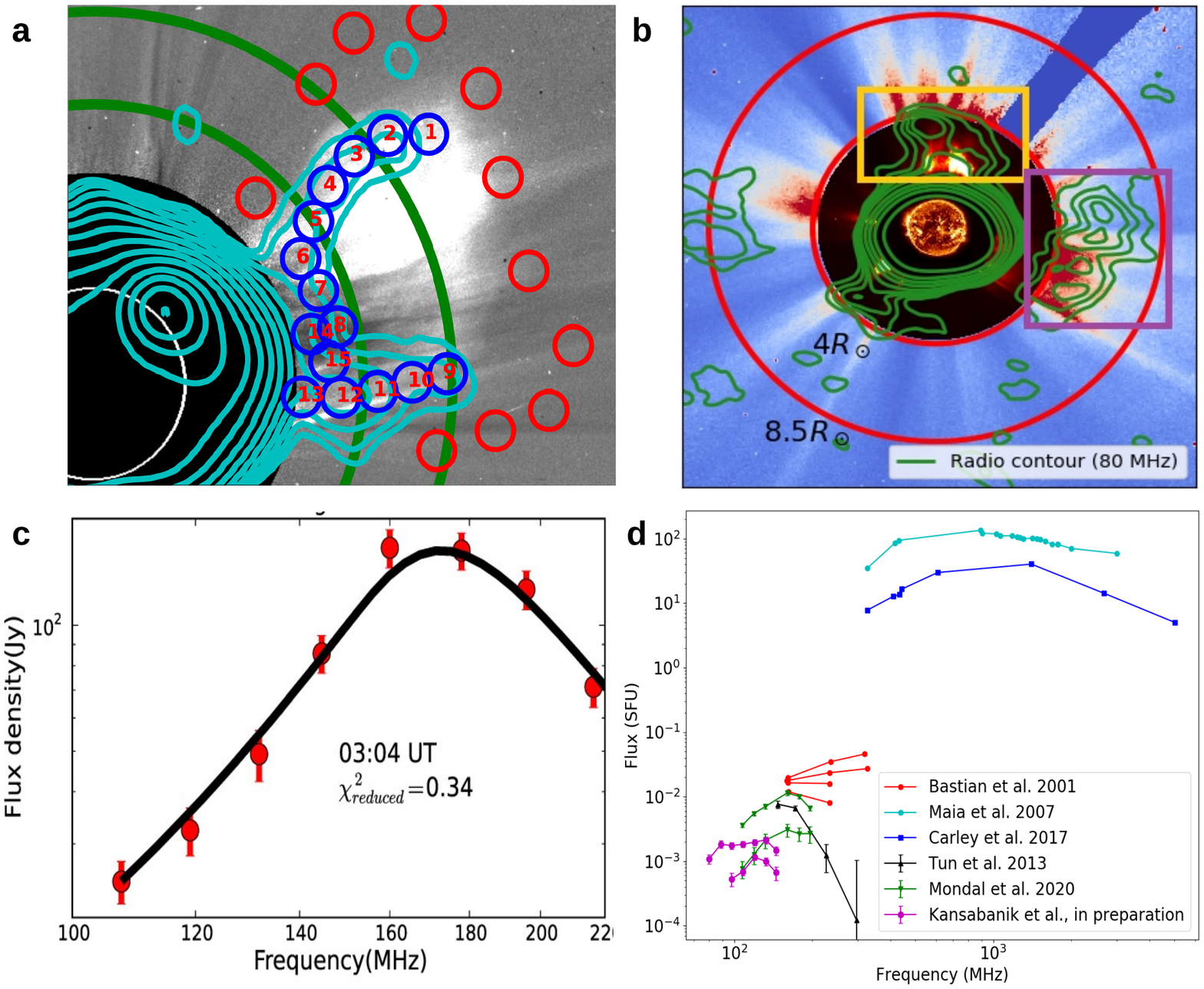}
    \caption{{\bf Radio emission from GS emission from CME plasma. }{\bf a. } Radio emission averaged over 108–145 $\mathrm{MHz}$ from the CME on 2015 November 4 is shown by the cyan contours and overlaid on a LASCO C2 difference image. Regions, where spectra have been extracted, are marked in blue circles. The green circles are drawn at 3 and 4 $\mathrm{R_\odot}$, respectively. {\bf b.} Radio emission from the CME on 2014 May 04 at 80 MHz is shown by the green contours overlaid on LASCO C2 and C3 white light images. Radio emissions from two CMEs are marked by yellow and purple boxes. The red circles are drawn at 4 and 8.5 $R_\odot$. {\bf c.} Observed spectrum from region 13 of the CME on 2015 November 4 at 03:04 UT is shown by the red points. The fitted GS spectrum is shown by the black solid line. {\bf d.} A comparison of the spectra from present works with the previous works is shown. Flux densities from the CMEs in the present work are much lower than detected earlier. (Panels (a) and (c) adapted from \citet{mondal2020a})}
    \label{fig:mondal2020_cme}
\end{figure*}

Coronal Mass Ejections (CMEs) are large eruptions of magnetized plasma from the solar atmosphere into the solar corona. These are the largest explosions in the Solar System and are the main drivers of space weather. Under the right conditions, CMEs can cause massive damage to satellites, lead to large-scale disruptions in power grids, and impact flight paths and GPS-based navigation systems among many other impacts on our technology-reliant society.
Hence understanding and predicting the effectiveness of a CME for causing space weather is globally a very active area of research. 
The extent to which a CME affects the Earth, or its geo-effectiveness, is primarily determined by the orientation of its magnetic field and measuring/ predicting this has been a challenge.
While several techniques have been used to estimate the magnetic field of CMEs, none of these are suitable for routine applications. 
Some of these well-known techniques are based on observations of -- splitting of type-II radio bursts observed in the dynamic spectra \citep[e.g.][]{Kumari2017,Mahrous2018}, circular polarisation of moving type-IV bursts \citep[e.g.][]{Raja2014,Kumari2017a}, and the standoff distance technique using EUV and white light images \citep[e.g.][]{Gopalswamy_2011,Schmidt2016}. 
Most of these methods are sensitive only to the average CME magnetic field, although, under some conditions, spatially resolved information about the magnetic field can also be obtained.

A promising approach for routinely measuring the CME magnetic fields is by modeling the GS emission from a CME. 
Measurements of GS emission can also be used to infer the distribution of nonthermal particles, another important contributor to space weather. 
This technique was first demonstrated by \citet{bastian2001}. 
Since CMEs are efficient particle accelerators, it is natural to expect that this emission should be there for all CMEs. 
Additionally, since this emission is expected to peak at the metric and decimetric wavelengths, low-frequency radio observations are very well suited for these measurements.
Despite significant effort being devoted towards it, this emission has been detected in only a handful of instances \citep{bastian2001,Maia2007,Tun2013,Bain2014,mondal2020a}.  

With the availability of high dynamic range images from AIRCARS and P-AIRCARS, it is now possible to detect much fainter GS emissions from CMEs. 
Figure \ref{fig:mondal2020_cme}d shows a comparison of all of the published gyronsynchrotron spectra with our work \citep[][and Kansabanik et al., 2022, in preparation]{mondal2020a}.
The flux densities detected in AIRCARS and P-AIRCARS images are much lower than those detected earlier. 
The radio emission from the CME analyzed by \citet{mondal2020a} is shown by the cyan contours overlaid on the LASCO C2 difference image of Fig. \ref{fig:mondal2020_cme}a. 
This was a rather slow and unremarkable CME and still, the authors were able to detect radio emission out to $\sim4.73\ \mathrm{R_\odot}$. 
For another event 
much fainter radio emission from two CMEs are detected simultaneously using P-AIRCARS (Fig. \ref{fig:mondal2020_cme}b). 
Here the CME GS emission is detected out to $8.3\ \mathrm{R_\odot}$, the largest heliocentric distance to date to which such emission has been detected.
This was also a slow CME. 
The recent detections using AIRCARS/P-AIRCARS imaging of MWA data suggest that the main reason behind the low historic success rate in the detection of CME GS emission has been the limitations imposed by the imaging quality available.

The MWA and AIRCARS/P-AIRCARS combination not only enables the detection of faint radio emissions from CMEs at large coronal heights, but the MWA data also provide a very good spectral sampling. 
This makes the fitting of the GS model to the observed spectra for estimating plasma parameters of the CMEs more robust. 
\citet{mondal2020a} extracted spatially resolved spectra from the multi-frequency observations covering $100-220\ \mathrm{MHz}$ at several PSF sized regions shown by the blue circles in the Fig. \ref{fig:mondal2020_cme}a. 
A sample spectra from Region 13 at 03:04 UT is shown in the Fig. \ref{fig:mondal2020_cme}c. Spectra are fitted using the numerical code for GS modeling by \citet{Fleishman2010}. 
The quality of the fit is self-evident. 
The estimated magnetic field strengths at different regions vary between $\sim7-13\ \mathrm{G}$, which are similar to those obtained by earlier studies. 

Although this method has proven to be a useful tool for remote estimating magnetic field and nonthermal electron distribution, the GS model has some degeneracies when used to estimate plasma parameters over a narrow frequency range. The most notable of these is between the estimated magnetic field strength and the angle it makes with the line of sight (LOS). 
The same spectral peak flux density can be achieved by multiple combinations of increasingly larger magnetic field strengths and smaller angles with the LOS.

A circular polarisation spectrum, if available, can break this degeneracy and this is a possibility we plan to explore using P-AIRCARS images.
Very recently our efforts have been rewarded by the very first detection of circularly polarised emission from a CME (Kansabanik et al. 2022, in preparation).
With a larger number of constraints on the GS model, enabled by polarimetric imaging, we can expect a more robust GS modeling of the spectra and a more accurate estimation of the physical parameters of CME plasma.
Our work provides evidence that the further increase in imaging quality and spectral coverage brought about by the availability of SKAO-Low will enable the GS modeling technique to be used for routine measurement of CME magnetic fields.
We envisage that this can turn SKAO-Low radio observations into a routine workhorse for space weather, much like the coronagraph measurements today.

\subsection{Understanding Coronal Heating}
\label{sec:WINQSEs}
The resolution of the so-called coronal heating problem is now generally understood to come from one or both of the following two processes: a) Alfv\'{e}n wave heating b) Nanoflare based heating \citep[e.g.][etc.]{moortel2015,klimchuk2015}. 
In the Alfv\'{e}n wave heating scenario, Alfv\'{e}n waves are generated in the convective zone and the photosphere due to the convective fluid motions, which then propagate to the corona where they eventually get dissipated through various pathways, depositing their heat into the corona. 
This scenario is believed to be efficient if the convective timescales are smaller than the Alfv\'{e}n timescale. 
When the Alfv\'{e}n timescale is larger than the convective timescale, the magnetic field lines get tangled and twisted, which later release this energy via magnetic reconnection processes, giving rise to small flares generally referred to as ''nanoflares". Though individually each nanoflare deposits a very small amount of energy, due to their large numbers, due to their large numbers they are believed to be able to deposit the energy at a rate sufficient to maintain the coronal temperature at an MK \citep{parker1988}.
As the convective deposition of energy into the coronal magnetic fields is active all over the Sun, Parker hypothesized that these small reconnection events must give rise to tiny flares, referred to as ``nanoflares", which must occur throughout the corona.

Nanoflares can accelerate electrons to supra-thermal energies, and when this beam of energetic electrons interacts with the thermal plasma it is passing through, it gives rise to coherent emission via plasma resonance processes \citep[e.g.][]{che2018}.
The coherent nature of this radio emission implies that for a flare of the same energy the observational signature at radio wavelengths is much stronger than the thermal emission signatures seen at EUV and X-ray bands.

This motivated some studies to detect the radio signatures of these nanoflares, but they were focused primarily on Type-I noise storms, which are generally associated with active regions \citep{mercier1997,ramesh2013}. 
However, for the nanoflare hypothesis to be true, such radio transients must also exist in the quiet corona. While this has been appreciated for a while, the available imaging quality has been insufficient to detect these low-contrast, rapid, narrow-band radio flashes.
\citet{mondal2020b} presented the very first evidence for the presence of ubiquitous impulsive narrow-band emissions from the quiet solar corona with flux densities of the order of a few mSFU, temporal widths generally smaller than 1s, and bandwidths smaller than 12 MHz. 
Based on their narrow-band and impulsive nature, they concluded that these emissions are nonthermal in origin and have been given the name of Weak Impulsive Narrowband Quiet Sun Emissions (WINQSEs). 
As the observed properties of WINQSEs meet all of the expectations for them to be the radio counterparts of nanoflares, this hypothesis was put forth by \citet{mondal2020b}.
Confirmation of this hypothesis requires significant additional evidence.

The first evidence towards testing this hypothesis comes from \citet{mondal2021a}, who identified the counterpart of a group of WINQSEs in the EUV band and then used it to estimate the energy associated with the WINQSEs group using the standard differential emission measure techniques. 
They found the energy associated with the WINQSEs group to be $\approx 3\times 10^{25}$ergs, suggesting that the energy associated with individual WINQSEs may likely lie in the nanoflare regime ($\sim 10^{24}$ ergs). 
A key necessary condition for this hypothesis to be true is that WINQSEs should always be present irrespective of the state of activity of the Sun. 
To investigate this, data from an extremely quiet solar condition
were analyzed and WINQSEs have been detected in this dataset as well (Mondal et al. 2022, submitted).
Additionally, since \citet{mondal2020b}, we have also improved our imaging and detection technique, and also fixed some issues which are generally unimportant for routine solar imaging, but need attention when dealing with flux densities of order a percent of the background. 
This forthcoming work will provide the first conclusive evidence for the ubiquity of WINQSEs during extremely quiet solar conditions.
Due to the significance of WINQSEs from a coronal heating perspective and also from the perspective of characterizing and understanding a new class of solar radio transients, work is ongoing on various fronts ranging from understanding their spectral and temporal characteristics. These include the use of machine learning algorithms for detection of WINQSEs and characterizing their morphologies (Bawaji et al. 2022, submitted), attempts to estimate the bandwidth of individual WINQSEs, exploration of their temporal profiles using high time resolution data, detection of EUV counterparts of WINQSEs to determine energies associated with these. 
The use of an independent imaging technique and pipeline using the so-called residual visibilities to look for WINQSEs-like features in the MWA data has also yielded consistent results \citep{sharma2022}.

It is useful to emphasize that with noise statistics of AIRCARS/P-AIRCARS images approaching thermal noise, we are already pushing the MWA data to its limits, especially in WINQSEs-related work. 
While incremental technical improvements will continue for some more time, they are not expected to lead to a large improvement in the imaging sensitivity or dynamic range of MWA images. 
We expect the SKAO-Low to be the instrument that will enable the next big step in this field and, hopefully, revolutionizes our understanding of this new class of radio transients called WINQSEs.

\subsection{Coronal Propagation Effects}
\label{sec:prop_effects}
\begin{figure*}[!t]
    \centering
    \begin{tabular}{cc}
    \includegraphics[trim={0cm 0cm 0cm 0cm},clip,scale=0.4,width=0.65\textwidth, height=0.2\textheight]{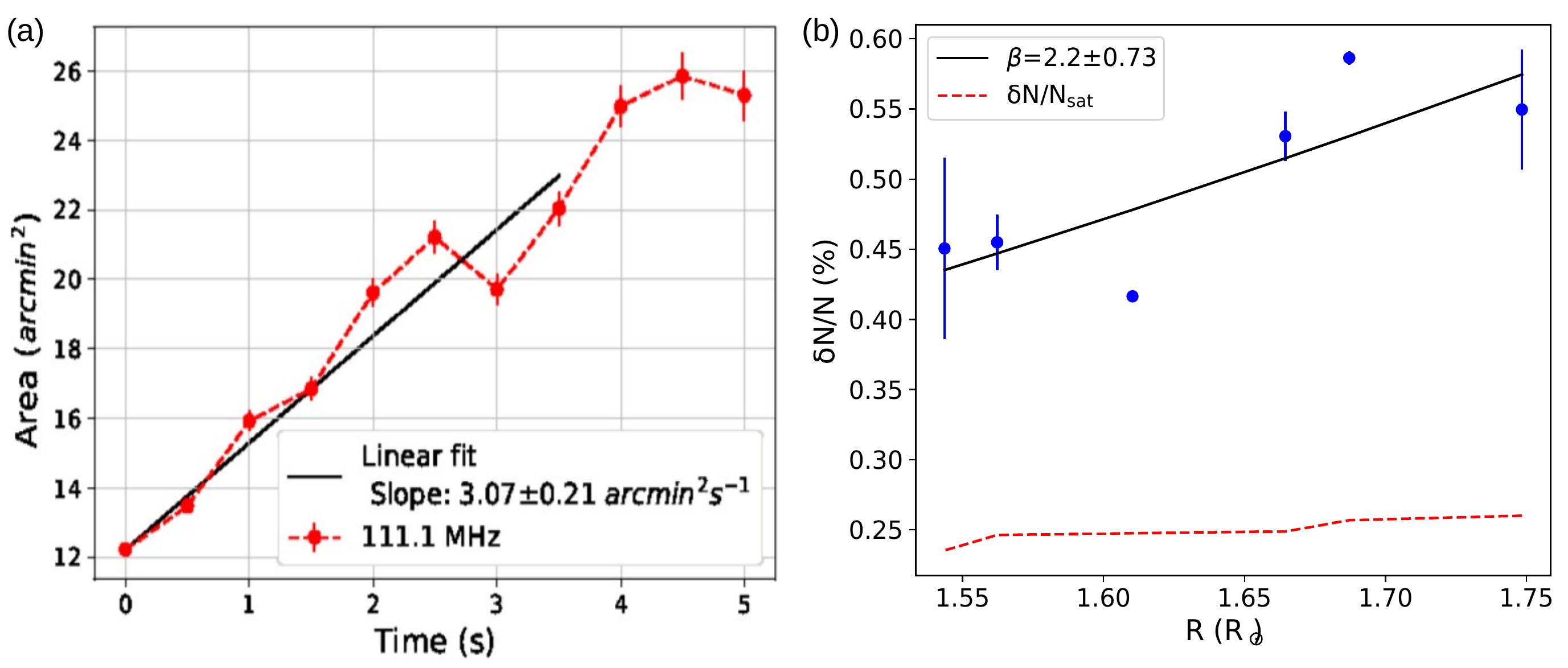} & 
    \includegraphics[trim={0cm 0cm 0cm 0cm},clip,scale=0.25]{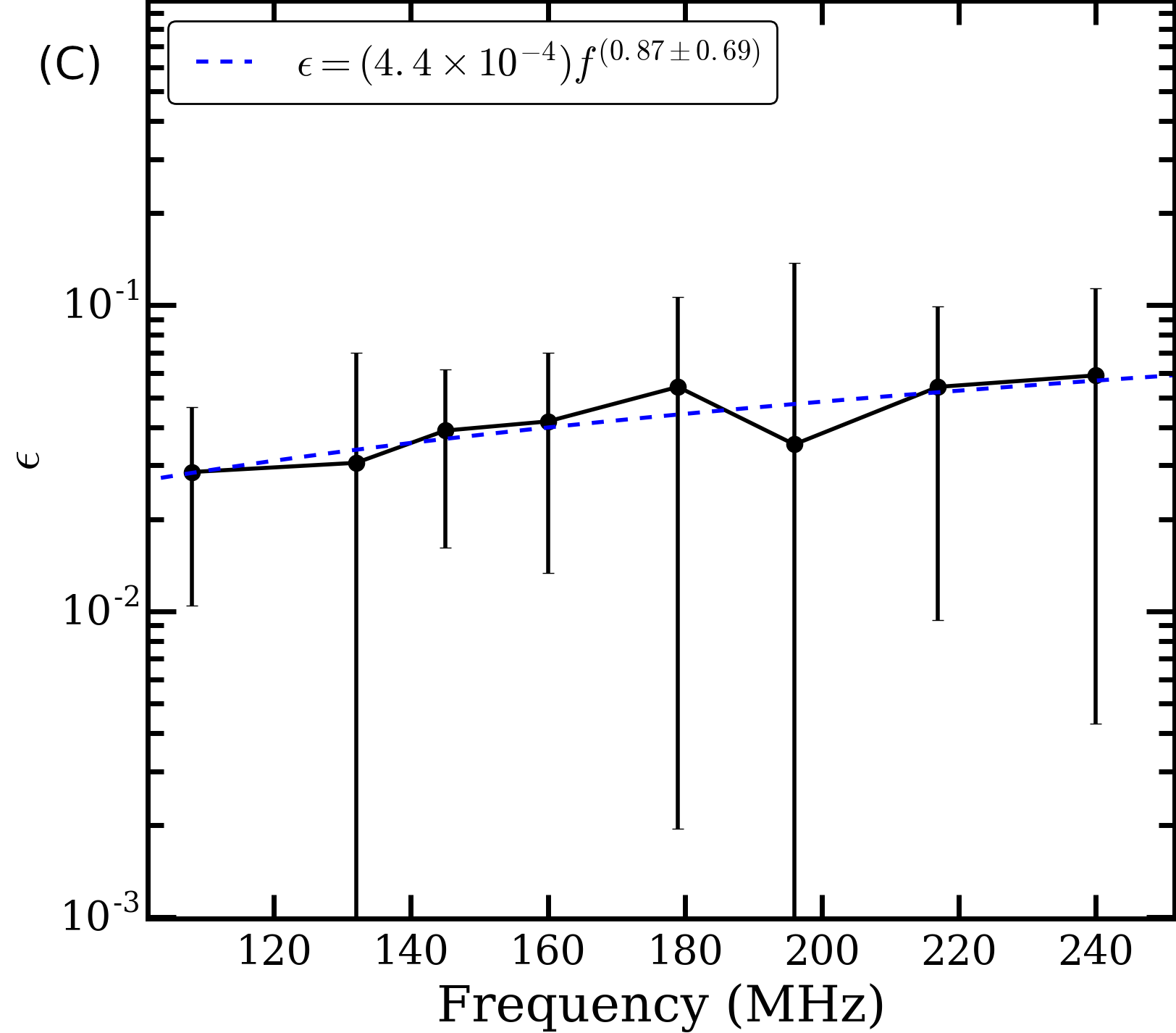} \\
    \end{tabular}
    \caption{{\it (a): }Evolution of the observed type-III source area. {it (b): }$\delta N/N$ derived using MWA data in 80 -- 200 MHz. The theoretical saturation limit ($\delta N/N_{sat}$) above which the medium is supposed to be in strong scattering regime is shown. (c) Variation of $\epsilon \sim \delta N/N$ at meterwave frequencies for the quiet Sun.}
    \label{fig:dnn}
\end{figure*}

With the typical inner coronal densities (heliocentric height $\lesssim$ 2$R_\odot$) varying in the range 10$^9$ -- 10$^8$\,$cm^{-3}$, the local plasma frequency ($\nu_p$=8.98$\sqrt{{n_e}}\,kHz$, where $n_e$ is the electron density in $cm^{-3}$) lies in the metrewave band. 
The refractive index in the medium, $n$, for radiation of frequency $\nu$ is given by
\begin{equation}
    n=\sqrt{1-\left(\frac{\nu_p}{\nu}\right)^2}.
    \label{eqn:ref_index}
\end{equation}
Being close to the local $\nu_p$, metrewaves are heavily influenced by the local density variations. 
These include the stochastic density fluctuations arising due to local turbulence as well as the spatial inhomogeneities in the density profile caused by plasma flows and magnetic field structures.
The effects of all these can be described under a common framework introduced by \cite{Arzner1999}. This was later revisited and improved and a publicly available modeling framework was developed by \cite{Kontar19_Arznercopy}. 

While analyzing the impact of the propagation effects at a frequency $\nu$, it is important to consider the variation in $\nu_p/\nu$ along the wave trajectory from the source to the observer. In the case of solar radio bursts involving coherent emission, like type Is and types IIIs, the emission originates in regions where $\nu \sim$ 1 -- 2$\nu_p$. 
Scattering is the dominant propagation effect that significantly alters the apparent location as well as the morphology of the source. 

\subsubsection{Propagation Effects under Quiet Sun conditions}
\label{subsec:stead-QS}
\begin{figure}[!t]
    \centering
    \includegraphics[trim={0cm 0cm 0cm 0cm},clip,scale=0.15]{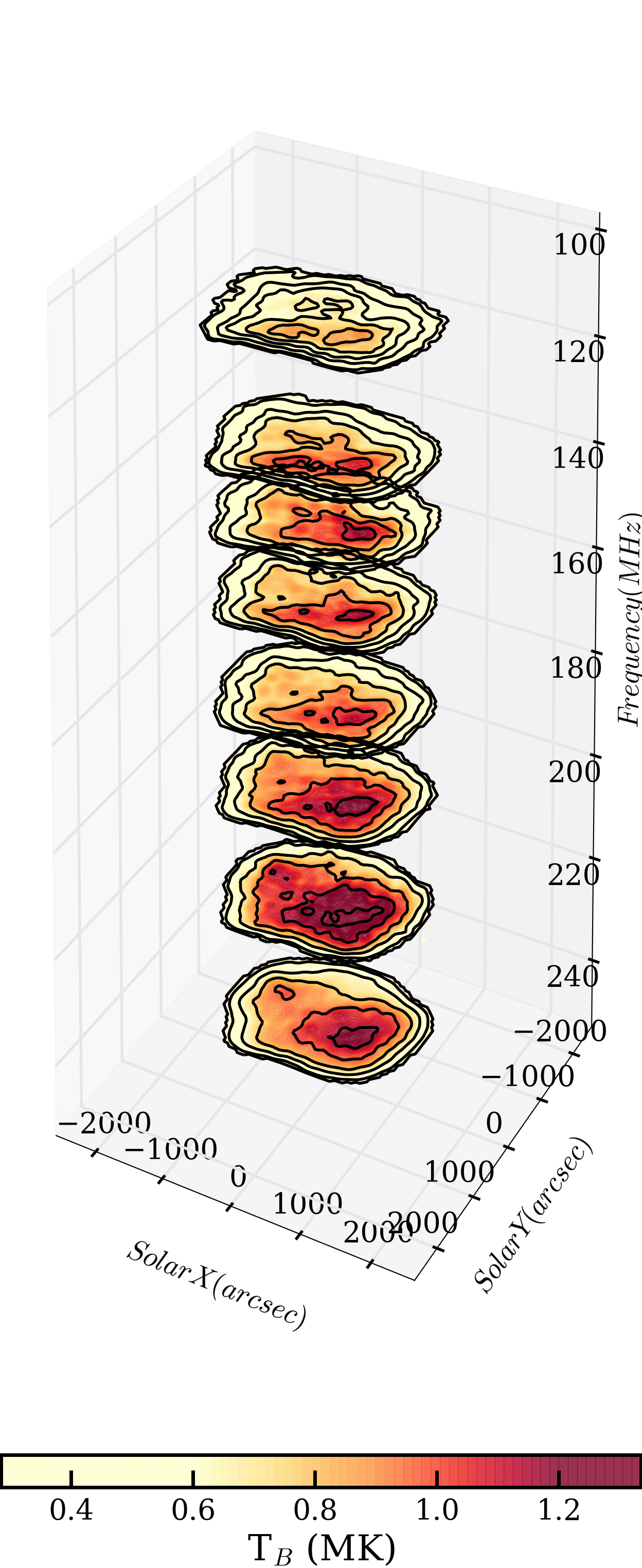}
    \caption{Quiet Sun morphology as a function frequency \citep{Sharma2020}. $T_B$ maps for 8 frequency bands from 109 MHz to 240 MHz. The frequency and time averaging are 2 MHz and 4 min respectively. The six contour levels corresponds to 10\%, 20\%, 40\%, 60\%, 80\% and 90\% w.r.t maximum $T_B$. (Reproduced from \citet{Sharma2020})}
    \label{fig:3D}
\end{figure}

High-fidelity studies of the quiet Sun at meter wavelengths are challenging due to a variety of reasons -- the presence of complex coronal magnetic field topologies; weak variations in the emission features superposed on the much brighter large angular scale emission, which is intrinsically harder to image. 
Till recently, it was common practice to make synthesis observations over many hours to get reasonable quiet Sun images at meter wavelengths \citep[e.g.][]{Kundu1987,Mercier2009,Vocks2018}. These studies established a relationship between optical, UV, and X-ray and meterwave emission features. 
For example in synoptic maps, optical wavelengths show large-scale counterparts in meterwaves \citep[e.g.][]{Kundu1987}. Features like coronal helmets and active streamers brighten meterwave thermal emission \citep[e.g.][]{Axisa1971}. Propagation effects like refraction and scattering induce an apparent shift in the observed locations of compact radio sources \citep[][e.g.]{Steinberg1971,Arzner1999,Gordovskyy2019,Kontar2019}. Scattering lowers the apparent coronal brightness temperature, especially at meterwaves \citep{Thejappa1994,Thejappa2008}.

The Figure \ref{fig:3D} from \citet{Sharma2020} presents a stack of $T_B$ maps as a function of frequency and provides a convenient way to track the evolution of the radio morphology of thermal bremsstrahlung emission with coronal height. 
They observed a gradual and systematic change in the single dominant compact emission feature in the western hemisphere, splitting into two segments at higher coronal elevations.
A key focus of this work was to quantify propagation effects using the novel approach based on a comparison of the simulated maps from a data-driven empirical model for thermal bremsstrahlung emission \citep[FORWARD;][]{Gibson2016}, which does not include the impact of propagation effects, with the observed solar maps made using MWA data.
At frequencies above 200 MHz, they found an excellent agreement in the modeled and observed solar flux densities integrated all over the solar disc (and hence independent of propagation effects like scattering), inspiring confidence in this approach. 
At lower frequencies, however, they found the observer flux density to be systematically lower than that from the FORWARD model.
Non-radial shifts as large as 8'-11' were found in the location of the peak of a bright active region, a clear indication of the presence of non-radial coronal structures. 
The angular size of the Sun was found to be 35–40\% larger in the MWA images.
They find scattering to play a dominant role in determining the observed $T_B$. As considerable radiation is scattered in and out of the lines of sight, the observed $T_B$ depends on both the intrinsic $T_B$ of the region from where the radiation is originating and the $T_B$ of
the neighboring regions. Their estimates of the level of density inhomogeneities lie in the range from 1--10\% and are consistent with earlier estimates.

\subsubsection{Propagation Effects on Burst Emissions}\label{sec:propeffect_bursts} 
The metrewave bursts, especially the type IIIs are excellent probes for studying the dynamics in the local coronal density structures. 
With the advent of modern interferometric arrays and snapshot spectroscopic imaging studies at sub-second and sub-MHz resolutions, investigations of stochastic properties of the coronal medium have now become tractable.
The presence of stochastic electron density fluctuations ($\delta N/N$, where $N$ represents the mean electron density and $\delta N$, the local departure from the mean) causes the radio waves from the burst source to undergo a diffusive transport resulting in a systematic growth in the apparent source size and broadening of the observed burst pulse. 
By tracking these effects systematically and modeling the source diffusion, \cite{mohan2019a} derived $\delta N/N$ in the inner corona below a heliocentric height of 2$R_\odot$.
Fig.~\ref{fig:dnn}(a) clearly shows the linear growth in the apparent source size.

\cite{mohan2021b} extended the analysis to a much larger height range using MWA data spanning 80 -- 200 MHz, and compared the results with other independent estimates of $\delta N/N$ from LOFAR in the 30 -- 80\,MHz band.
Figure~\ref{fig:dnn}(b) shows the variation of $\delta N/N$ with height derived by \cite{mohan2021b} with dots and the best fit powerlaw function with spectral index, $\beta \sim $2.2$\pm$0.73. 
Similar to \citet{mohan2019a}, this study also found the $\delta N/N$ values to be higher than the saturation limit, $\delta N/N_{sat}$, estimated using the model by \cite{Arzner1999}, or that one is in the so-called strong scattering regime where wave propagation is well described by a pure diffusion framework. 
This implies that the other sources of propagation effects like refraction, radio wave ducting due to randomly oriented fibrillar over-dense structures \citep{robinson_scat1983} and radio echos \citep{Kuznetsov20_radioEcho_driftpair} from random over dense pockets in the medium, which could give rise to non-diffusive effects in the observed evolutionary profiles, can safely be ignored. 
The study revealed that the variation in $\delta N/N$ across the inner corona height range of 1.4 -- 2.2 $R_\odot$ is within a factor of few. 
Since weak type-III bursts are associated with weak solar flares which are quite frequent and associated with active regions with widely varying physical conditions, more studies along these lines can help develop a more detailed picture of the characteristics and evolution of density fluctuations in the inner corona.

\subsection{Polarisation Studies of Solar Radio Emissions} 
Polarisation studies of the radio emissions (both thermal and nonthermal) from the solar corona are sensitive probes of the emission mechanisms involved and the coronal magnetic fields - the hard-to-measure driver of coronal dynamics and space weather.
In broad terms, the anisotropies in the coronal medium due to the all-pervading coronal magnetic field leave their imprint on the polarisation properties of the radiation originating from this medium.
While their scientific merits have been appreciated for a while, high-fidelity polarimetric radio imaging poses significant technical challenges.
For this reason, most of the polarimetric studies of the solar corona thus far have relied on non-imaging dynamic spectrum measurements.
Some instruments have also used simultaneous Stokes I imaging with Stokes V dynamic spectra to help with the localization of the source of polarised emission \citep{Sasi2014PhDT}.

As discussed in Sec. \ref{sec:P-AIRCARS}, we have recently developed a full Stokes snapshot spectroscopic imaging pipeline optimized for solar imaging with the MWA \citep{Kansabanik2022b, Kansabanik2022c, Kansabanik2022d}. 
It offers polarimetric calibration on par with usual interferometric imaging and we expected it to enable multiple new avenues for exploration with the MWA and the future SKAO when it becomes available. 
A limited but illustrative set of examples of the kinds of studies polarimetric imaging capability will enable include the following --
It has been reported that thermal emission (for instance from the coronal streamers) split into two modes o-mode and x-mode due to radio wave propagation in the solar corona \citep{2010ApJ...711.1029R}. It is known that o-mode and x-mode are circularly polarized with opposite senses of polarisation. 
Polarimetric imaging capability will enable measurements of the degree and sense of circular polarisations and their energy budgets \citep{2010ApJ...719L..41R, Ram2013}.
Detailed polarimetric studies of well-known solar bursts, including identifying if the emission comes from Fundamental or Harmonic based on their degree of polarisation \citep[e.g.][]{Ram2011,Sas2013}.
Routine well-constrained modeling of gyrosynchrotron emission from CMEs (Sec. \ref{sec:CME-gyro}).
Examining and understanding the polarimetric properties of the newly discovered WINQSEs (Sec. \ref{sec:WINQSEs}).
Estimation of mean coronal magnetic field for the quiet Sun regions using the low level ($<$1\%) circular polarisation arising due to the birefringence of the coronal medium \citep{Sastry_2009}.
In addition, as this would be the first exploration of this phase of space, it also offers considerable discovery potential.

\section{Heliospheric Physics}
Heliospheric physics deals with exploring and understanding the plasma of solar origin permeating the entire solar system.
This medium carries the influence of the Sun to the Earth (and all other planets) and this is where Space Weather develops and takes place.
The vastness of this medium implies that space-based probes which provide local measurements can only offer a very limited and insufficient sampling.
The tenuousness of this medium makes it exceptionally hard to study, though in the past few decades there have been some heliospheric imagers that have extended the idea of coronagraphs to the entire heliosphere with remarkably success \citep{Jackson-SMEI-2004, Eyles-STEREO-HI}. 
Well before these instruments became available, radio observations in the metrewave band had already established themselves as efficient remote sensing probes of the interplanetary medium by studying the propagation effects on emissions from celestial sources and spacecraft radio beacons. 
These include angular and spectral broadening \citep{WoA95}, Faraday rotation \citep{Bird-CME-FR-1985} and phase and amplitude scintillations due to the interplanetary medium  \citep{HSW64, Kojima-IPS-review-1990}.
The following sections discuss the various investigations of the heliosphere enabled by radio observations using existing and the future SKAO-Low.

\subsection{Interplanetary Scintillation}
\label{sec:IPS}
The outermost layer of the Sun, the so-called corona, has a very high temperature ($\sim{10^{6}}\,$K) causing a huge pressure difference between the corona and the interplanetary medium. The high-pressure gradient, despite the restoring influence of solar gravity, causes the coronal plasma to escape out into the interplanetary medium (IPM), and eventually form the solar wind. The solar wind due to its expansion through the interplanetary space develops fluctuations \citep{Par58,Par63} or turbulence \citep{BCa05} in velocity, magnetic field, and density. The scale sizes of such fluctuations often show a wide range of spatial scales.
The large-scale sizes in the solar wind, above 10${^4}$ km, generally, follow a power-law spectrum \citep{IWo70}.  
Fluctuations of scale size greater than 10${^7}$ km also occur in the solar wind.  
These develop principally due to stream structures in the solar wind, and cannot be explained by the turbulence spectrum. Unlike velocity and magnetic field fluctuations, density fluctuations are ubiquitous in the solar wind.  Further, density fluctuations are also believed to be a better tracer of solar wind flows as compared to the background solar wind density, N \citep{ACK80,WoA95,HWN95}. Hence, knowing how density fluctuations ($\Delta$N) or relative density fluctuations ($\frac{\Delta{N}}{N}$) vary in time and with distance from the Sun is crucially important for a variety of 
applications ranging from understanding turbulent dissipation to the consequent local heating in the solar wind.

\begin{figure}[ht]
\centering
\includegraphics[width=9.0cm,height=7.0cm]{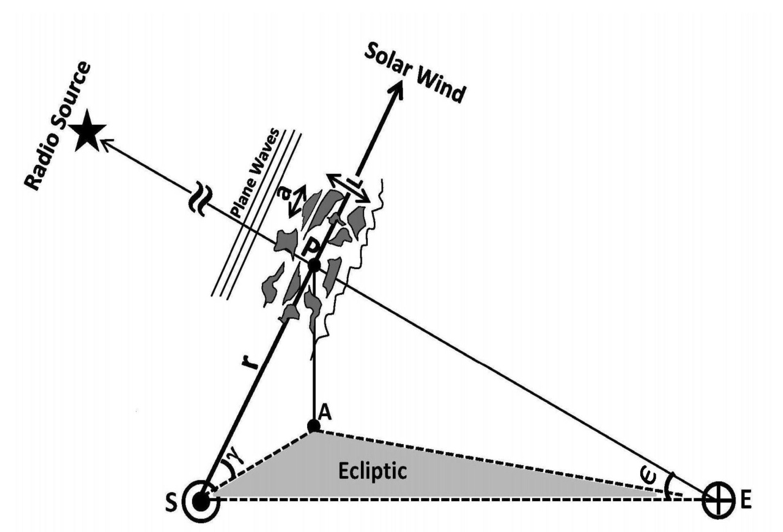}
\caption{\textbf{A schematic of the IPS observing geometry.} A schematic of the IPS observing geometry.  The Earth, the Sun, the point of closest approach of the LOS to the Sun (P), and the foot point of a perpendicular from P to the ecliptic plane are shown by points E, S, P, and A respectively, while the angles 
$\epsilon$ and $\gamma$ are the solar elongations and the heliographic latitude of the observed source.  The radial distance from the Sun, 'r' of the LOS to the radio source is generally given by sin($\epsilon$). (reproduced from \cite{BiJ14})}
\label{fig1-ips}
\vspace{-0.5cm}
\end{figure}

The variations in solar wind density fluctuations can be measured through a phenomenon known as ``Interplanetary scintillation (IPS)", which was first reported by \cite{HSW64}. 
IPS is the radio analog of the optical twinkling of stars which is caused by turbulent density fluctuations in Earth's atmosphere.
In IPS the radio waves from distant point-like extra-galactic radio sources 
are scattered by density fluctuations in the solar wind and give rise to amplitude fluctuations in the radiation incident on ground-based radio telescopes.  
The typical geometry during IPS observation is shown in Fig. \ref{fig1-ips}. The angle $\epsilon$ is the solar elongation while `r' is the heliocentric distance of the radio source, in AU, and is given by r = sin($\epsilon$).  The scintillation measured at 327 MHz at the Earth are modulated by a Fresnel filter function $\mathsf{Sin^{2}(\frac{q^{2}\lambda z}{4\pi})}$, where q is the wave number of the density irregularities, z is the distance from E to P, and $\lambda$ is the observing wavelength. It must be noted that IPS at 327 MHz are caused primarily by solar wind density fluctuations with scale sizes $\leq$10${^3}$ km. 
\citep{RBA74,CFi85,YaT98,FBD08}. 
On the other hand, due to the action of the Fresnel filter, scale sizes $\geq$ $10^{3}$ km result in scintillation only at distances $>$ 1 AU. In other words, the Earth is well within the Fresnel or the near zone for these scale sizes. The IPS phenomena thus acts as an in-built filter that doesn't allow contributions to the measured scintillation at the Earth from large-scale size density fluctuations. 

IPS observations have been used to study the fine scales structures in cometary ion tails as they occult well known IPS sources.
\citep{ARB75,SlM87,JaA91,JAl92,JaA92}. 
It is found that a regime dominated by high values of density fluctuations at small spatial scales exist along and close to the axis of cometary plasma tails, which is usually an order of magnitude greater than that in solar wind. 
On the other hand, a fine scale regime, similar to that of the solar wind plasma, also exists close to the edge in cometary tail plasma. Both the regimes have the ability to produce enhanced scintillation as observed through IPS observations for Comets Austin and Halley \citep{JaA92}.

IPS measurements along the line of sight to the radio source can be quantified using the scintillation index (m), which is defined as the ratio of the scintillating flux ${\Delta}S$ to the mean source flux $<$S$>$, $\mathsf{m = \frac{{\Delta}S}{<S>}}$.  
Thus, an ideal point source, for which the scintillating flux is equal to the mean source flux, will have a $m$, of unity at some fixed distance from the Sun. In fact $m$ increases with decreasing $r$ or $\epsilon$, until a certain $\epsilon$.  
This region is known as the weak scattering region. Beyond this point, $m$ falls off sharply as it approaches the near-Sun region known as the strong scattering region.  The $\epsilon$ at which $m$ turns over is dependent on frequency.
At 327 MHz it lies between 10${^\circ}$ and 14${^\circ}$ or $\sim$40 R${_{\odot }}$, where, R${_{\odot}}$ is the solar radius. In the weak scattering region, the approximation of scattering by a thin screen is valid \citep{Sal67}, which means that most of the contribution to the scintillation will come from the point ``P" on the LOS that is closest to the Sun. Plane waves from distant, compact extragalactic radio sources on passing through the thin screen of density fluctuations will develop rms phase deviation, $\phi{_{rms}}$, given by 
\begin{equation}
\centering
\phi_{rms} =(2\pi)^{\frac{1}{4}}{\lambda}r_{e}(aL)^{\frac{1}{2}}[<{{\Delta}N}^{2}>]^{\frac{1}{2}},
\end{equation}
where $r_{e}$ is the classical electron radius, $\lambda$ is the observing wavelength, and $a$ is the typical scale size in the thin screen of thickness $L$. 
In the weak scattering regime, $m = \sqrt{2}{\phi{_{rms}}}$, and thus using the variations in $m$, the changes in solar wind density fluctuations can be inferred. The measurements of $m$ have been made possible through ground-based IPS observations of a single station facility at Ooty Radio Telescope (ORT) in India since the 1970s and a three-station facility at the Institute for Space-Earth Environmental Research (ISEE) in Japan since the 1980s. There are other ground-based IPS facilities such as MEXART in Mexico \citep{MEXART2010, MEXART2016} and Pushchino in Russia, providing daily routine measurements for $m$. 
The IPS studies reported here use observations from the ORT and ISEE facilities.

In the past few decades increasingly sophisticated algorithms, using IPS observations as their primary observables, have been developed for 3D reconstruction of  global heliospheric parameters -- density and velocity \citep[e.g.][]{Jackson1997, Jackson1998}. 
More recently these reconstructions have been extended to include data IPS stations from across the world, from Thompson scattering based heliospheric imagers and also incorporate MHD models.
These models can now also provide a reconstruction of the heliospheric magnetic fields \citep{Jackson2020} and have recently been used for a detailed modeling exercise for a CME \citep{Iwai2022}. 

In the past few years, the wide FoV imaging capability of the MWA has also been put to use for IPS observations \citep{MWA-IPS-1-2018}. 
A novel aspect of these studies is that they use high time resolution imaging to simultaneously observe a large number of IPS sources distributed across the large MWA FoV and represents a significant advance in the field.
Other interesting aspects of these studies are that, in addition to be being used for heliospheric physics, these observations are also being used to address other science objectives ranging from understanding the properties of these very compact sources at low radio frequencies \citep{MWA-IPS-2-2018, MWA-IPS-3-2018} to carrying out an all sky survey of compact radio sources \citep{MWA-IPS-5-2019}.
Very recently IPS observations have also been demonstrated using the Australian Square Kilometre Array Pathfinder \citep[ASKAP;][]{ASKAP-IPS-1-2022} -- another wide FoV imaging instrument, but operating at frequencies much higher than the MWA and other instruments used for IPS.
While limited in scope, this is a very interesting development. 
The higher observing frequencies of ASKAP allow one to use IPS to probe regions of the heliosphere at much smaller elongations than possible with most IPS instruments.

\subsection{Determining Angular Diameter of Radio Sources}
IPS has been demonstrated to be a very efficient way to detect and characterize sub-arcsecond structures in radio sources {\citep{LHe66,LHe68}} and a recent example is \citet{MWA-IPS-2-2018}.
Assuming a Gaussian brightness distribution for the compact component of the radio source, the systematic variation of $m$ with $\epsilon$ in the weak scattering region can be used as an indicator of angular source size of scintillating radio source \citep{JAl93}. 
An independent method of estimating angular diameter is based on fitting the observed IPS power spectra with solar wind parameters including the radial speeds, the strength of scintillation, inner and outer scales, power law index of density fluctuations along with the angular size of a scintillating component of the radio sources \citep{MAn90}.
\subsection{Variations in Large-scale Structure of Solar Wind}
Since IPS observations are primarily sensitive to density fluctuations of scale sizes less than 1000 km (typically at a frequency of 327 MHz), they 
can probe a large region of the inner heliosphere, covering from 0.2 to 0.8 AU, and have been successfully used to study the large-scale structure of solar wind (of scale sizes $\approx$ 100--1000 km) as a function of helio-latitudes and heliocentric distances over several solar cycles \citep{ACK80,JaB96,Man12}.
\subsection{Probing Density Fluctuations in Interplanetary Medium}
It is important to note that IPS measures only the fluctuations, and not the bulk density itself.  However, any enhancement or decrease in IPS level in the solar wind is associated with corresponding variations in density. Thus, whenever interplanetary disturbances travel in the IPM, 
containing either enhanced or depleted rms density fluctuations as compared to the background solar wind, they exhibit themselves as changes in the level of scintillation. 
It implies that whenever solar wind turbulence levels change, they will be reflected in IPS measurements as changes in $m$.  It is worth mentioning here that IPS is sufficiently sensitive to changes in $\Delta{N}$ that it has been used to probe density fluctuations in tenuous cometary ion tails \citep{JaA92} and to study solar wind disappearance events wherein average densities at 1 AU drop to values below 0.1 $cm^{-3}$ \citep{JaF05}. 
On the other hand, IPS measurements have also been useful to study the propagation of large-scale disturbances like CME \citep[e.g.][]{Man06}. The CME while propagating in the IPM causes an increase in density and turbulence in its sheath region, which can be mapped by obtaining the enhancement in the g-index (g = ${\Delta}S/{\overline{\Delta{S}}}$) for each observation of all the radio sources obscured by the CME. The g-maps, thus produced, show enhanced regions of density and turbulence where g $>$1. Further, using time series of IPS g-index and velocity and time-dependent computer-assisted tomographic technique, the three-dimensional heliospheric reconstruction of solar wind velocity and density have also been obtained, which help in tracking the traveling interplanetary disturbances associated with CMEs \citep[e.g.][]{Man10}.

\subsection{Probing Long-term Variations of Solar Cycle Magnetic Activity}
\begin{figure}[ht]
\centering
\includegraphics[width=8.0cm,height=6.0cm]{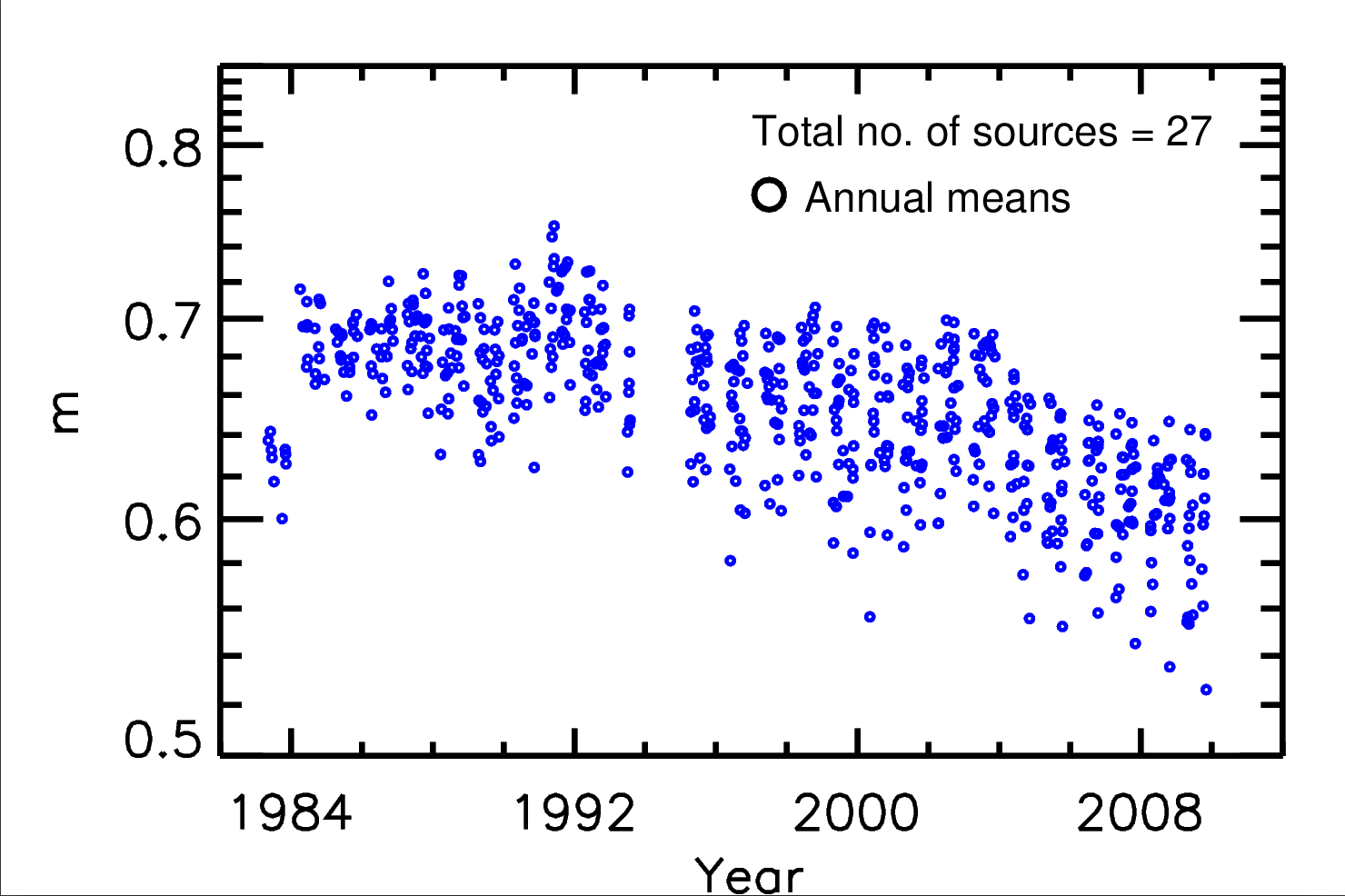}
\caption{\textbf{Annual variation of scintillation index level.} Shows the annual means of $m$ as a function of time, 
covering the period between 1983\,--\,2009, for a total of 27 IPS radio sources (reproduced from \cite{JaB11}).}
\label{fig2-ips}
\end{figure}

The study by \citet{JBG10} of solar cycle changes of solar photospheric fields using ground-based magnetograms has shown that there has been a steady decline in their strength at high latitudes ($>$45$^{\circ}$) since 
$\sim$mid-1990's.  Since solar high-latitude magnetic fields supply most of the heliospheric magnetic flux during solar minimum conditions \citep{SCK05}, the long-term decline in solar field strength implies that it would affect the heliospheric open flux and, in turn, the magnetic field fluctuations or the solar wind turbulence in the fast solar wind at high-latitudes. A steady decline in coronal magnetic field strength correlated with the reduction in photospheric field strength has also been noticed \citep{SaJ19}. Using IPS measurements of $m$ at 327 MHz obtained from the ISEE facility between 1983 and 2009, \cite{JaB11} reported the temporal variations in $m$ during solar cycles 22--23. For the 27 selected IPS sources, including a point source 1148-001, which had at least 400 
individual observations over this period of 27 years, the values of $m$ were made distance independent by normalizing every individual observation for each source by the value of $m$ for the source 1148-001 at the corresponding $\epsilon$. The temporal variations obtained thus for the normalized $m$ for all of the 27 sources, as shown in Fig. \ref{fig2-ips} that depicts the annual means of $m$ for each source in blue open circles as a function of time, unambiguously showed a steady decline in $m$ starting from around mid-1990s. The reduction in $m$ implies a global 
reduction in solar wind turbulence levels as expected. It has been further noticed \citep{IJB19} that the steady decline of solar wind turbulence levels has been continuing until the conclusion of the solar cycle 24, in sync with the continuing decline in high-latitude photospheric field strength.

\subsection{Locating Signatures of Unusual Polar Reversal}
The polarity of the Sun's magnetic field in each solar hemisphere reverses around the maximum of each 11-year solar cycle, the well known polar reversal. 
The occurrence of the reversal process ultimately decides 
the strength of the polar field at the end of a solar cycle, which acts as a precursor for predicting the amplitude of the upcoming solar cycle. The signatures of the polar reversal can be identified in the solar wind as polar coronal holes. 
These coronal holes are the source regions of high-speed solar wind that begin to develop in both solar hemispheres after the field reversal takes place \citep{KTR73,NoK76,Zir77}. Further, it is known that polar coronal holes usually fully developed $\sim$2 years after the field reversal \citep{FuT16}. 
In cycle 24, the polar reversal showed an unusual pattern in the photospheric synoptic magnetic maps, wherein the southern hemisphere underwent a single and clean polar reversal in Nov 2013 while the polar reversal in the northern hemisphere started in Jun 2012 and was completed after more than a year, in Nov 2014 \citep{JaF18}. Signatures of this unusual field reversal pattern were identifiable in the solar wind velocity maps, produced using a tomographic technique applied to multi-station IPS observations at 327 MHz obtained from the ISEE facility, which showed well-defined polar coronal holes in the southern hemisphere just around 2 years after the field reversal while the polar coronal holes were not fully developed in the northern hemisphere. 
A difference in time of appearance of fully developed polar coronal holes in the two hemispheres thus clearly indicates the completion of reversal occurred at different times in the two hemispheres.

\subsection{Probing Solar Wind Turbulence and Magnetic Field Fluctuations}
\begin{figure}[ht]
\centering
\includegraphics[width=8.0cm,height=6.0cm]{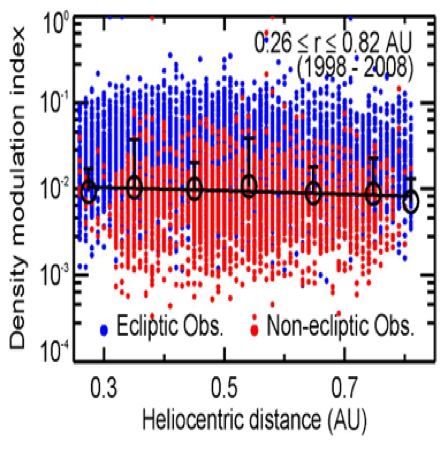}
\caption{Spatial variation of the density modulation index, ${\epsilon}_N$, of all 27 selected sources in the period from 1998 to 2008. While the blue and red solid dots are the actual measurements of normalized modulation indices for ecliptic sources and non-ecliptic sources, respectively, the large open circles in black represent averages of all 
observation at intervals of 0.1 AU. The solid line is a fit to these average values \citep[reproduced from][]{BiJ14}.}
\label{fig3-ips}
\end{figure}

In another study of the inner heliospheric solar wind during the solar cycle 23 and covering heliocentric distances of 0.26 - 0.82 AU, the variations in relative density fluctuations or solar wind density modulation index defined as, ${\epsilon}_{N}$ = $\frac{{\Delta}N}{N}$, were computed \citep{BiJ14}.  
The values of ${\Delta}N$ at a particular heliocentric distance were directly computed using measurements of $m$ from ISEE at that distance, while values of $N$ at 1 AU were extrapolated using a solar wind density model to determine the value of $N$ at the heliocentric distance where the ${\Delta}N$ values were estimated.  Also, the values of $m$ were made source size independent before computation of ${\Delta}N$. It was found that the variation of ${\epsilon}_{N}$ both for the ecliptic and non-ecliptic radio sources was roughly constant with heliocentric distance as depicted 
in Fig. \ref{fig3-ips}.  It is shown that the length scales probed by the IPS observations at 327 MHz are in the inertial range of the turbulence spectrum. Thus, it is reasonable to presume that the magnetic field is frozen in the solar wind plasma and the density fluctuations can 
be taken as a proxy for magnetic field fluctuations \citep[e.g.][]{Spa02}. Furthermore, it is found that the temporal variation of ${\epsilon}_{N}$, after making the ${\Delta}N$ values distance independent, showed a decline of around 8\% during the period 1998--2008. Presuming a linear relationship between the relative density fluctuations and the magnetic field fluctuations \citep{SSp04}, it appears reasonable that the magnetic field fluctuations have also been declining steadily over the same period. The decline in solar wind density fluctuations as inferred by IPS observations is, therefore, claimed to be linked to the steady decline of solar photospheric magnetic activity.

\subsection{Tracking Source Region of Geomagnetic Events}
Over the years, it has been established that when the interplanetary magnetic field (IMF-$B_{z}$) becomes southward, it produces a strong coupling between the solar wind and the Earth's magnetosphere through magnetic reconnection of the IMF-$B_{z}$ and the terrestrial magnetic field, leading to geomagnetic storm events \citep{Dun61}. 
Further, it is known that the long-duration southward IMF-$B_{z}$ events (magnitudes greater than -10 nT and lasting for over 3 h) play a significant role in triggering magnetic storms 
\citep{GTs87,TsL92}. Tracking back the source regions of such geomagnetic events, which are, often, associated with well-defined solar eruptive phenomena such as CMEs is crucial as their interplanetary counterparts preferentially result in southward IMF-$B_{z}$ condition at 1 AU. The synoptic velocity maps, those produced using IPS observations via tomographic technique and then projected back to the source surface, at 2.5 R${_{\odot }}$, can be used to track the source region of the southward IMF-$B_{z}$ events. From the tomographic velocity map, the velocity of solar wind flow in the traceback region can be identified and can be linked to the source region. \cite{BiC16}, using the IPS tomographic maps, tracked the source region of a prolonged southward IMF-$B_{z}$ that produced severe geomagnetic storm conditions,  and were able to locate the high-velocity solar wind flows in the traceback region, which had occurred because of the eruption of a CME on the Sun.
\subsection{IPS with the SKAO}
The high sensitivity observations of a large number of IPS radio sources per day can be carried out using the SKAO at different frequencies as compared to the current dedicated ground-based IPS facilities such as ORT and ISEE etc. as well as radio interferometers such as LOFAR and MWA etc., which can provide IPS observations between $\sim$ 200--1000 IPS radio sources per day. Through IPS observations of a large number of radio sources along the multiple LOS, complete distribution of solar wind plasma properties such as density and velocity variations those sensitive to IPS, can be obtained. This, in turn, can be employed to estimate 
the systematic variation in the scintillation index as a function of heliocentric distance and heliolatitudes. In addition, they can be used to produce a three-dimensional heliospheric reconstruction of traveling disturbances in the solar wind such as CME and so on, providing a more robust estimation of solar wind density and velocity, as compared to the other IPS observations mentioned earlier. The high sensitivity IPS observations thus obtained from the SKAO shall be used

\begin{enumerate}  
\item for studying variations in solar wind plasma, especially those of density fluctuations, as 
discussed earlier.
\item for studying large-scale structure of solar wind and its variation as a function of heliocentric distance 
and heliolatitudes.
\item for studying long-term solar wind variations such as solar cycle changes of solar wind associated 
with photospheric magnetic activity changes.
\item for tracking CME propagation as well as tracking solar wind disturbances traveling in IPM along 
with coronal white-light images obtained from other spacecraft such as ADITYA-L1, Parker's 
Solar Probe and Solar Orbiter etc.
\item for the heliospheric reconstruction of solar wind density, velocity, and IMF-$B_{z}$ at 1 AU to make IPS-based prediction of space weather events using the aforementioned parameters. 
\item building a robust IPS-based estimated system for predicting the arrival of CME at 1 AU using 
global MHD simulations along with white-light coronal observations.
\end{enumerate}
\subsection{Heliospheric Faraday Rotation} \label{sec:helio-FR}

\begin{figure*}[!t]
\centering
\includegraphics[trim={1.5cm 4.1cm 1cm 2.7cm},clip,scale=0.65]{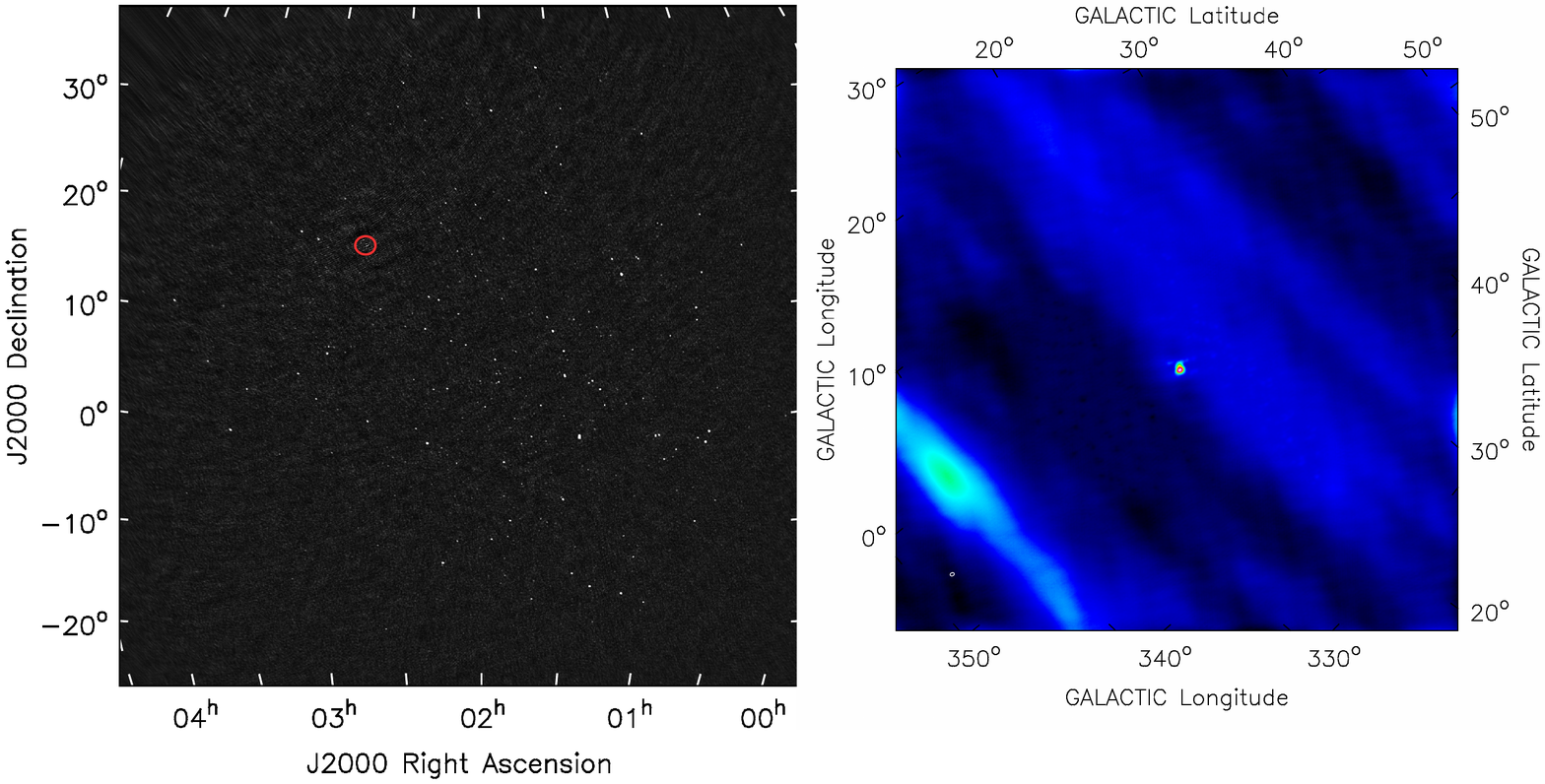}
\caption{{\bf Stokes I emissions from the background radio sources and galactic plane in the presence of the Sun in the FoV.} {\bf Left panel: } Image centered at 80 $\mathrm{MHz}$, $\sim 60^{\circ}$ on each side, obtained over 2 minutes and 2.28 $\mathrm{MHz}$ after subtraction of modeled solar visibilities from the data. The image is from the observation on 2014 May 04 using the MWA, and the red circle with radius 2 $\mathrm{R_\odot}$ is the region where the Sun was present. (reproduced from \citet{Kansabanik2022a}). {\bf Right panel: } The Sun is present at the center of the image. The extended emission at the bottom left of the figure is the galactic diffuse emission from the galactic plane. (Left panel reproduced from \citet{Kansabanik2022a})}
\label{fig:other_sources}
\end{figure*}

It is well established that solar magnetic fields are a key driver of all kinds of space weather phenomena. Their routine measurements are only available at the solar photosphere and from in-situ spacecraft measurements usually near the Earth or at certain vantage points in the heliosphere. 
There are other techniques available to measure magnetic fields up to few tens of solar radii using observations at different wavebands \citep[e.g.][etc.]{Gopalswamy_2011,Kumari2017,Ramesh2021,Sas2013,bastian2001,mondal2020a, Sasi2022a}. Measurements of the magnetic fields in the solar wind and CMEs over the entire heliosphere remain a challenge and are only rarely possible. 

When linearly polarised radiation passes through a magnetized plasma, its plane of polarisation rotates. The magnitude of Faraday rotation (FR) experienced depends on the three quantities; i) electron density ($\mathrm{n_e}(\vec{r})$), ii) magnetic field ($\mathrm{\vec{B}}(\vec{r})$) and the wavelength of the emission ($\lambda$), and is given by
\begin{equation}
\begin{split}
    \mathrm{FR}&=\lambda^2\times\mathrm{\frac{e^3}{8\pi \epsilon m_e^2c^3}\int_{obs}^{src}\mathrm{n_e}(\vec{r})\mathrm{\vec{B}}(\vec{r})\ .\ d\vec{s}}=\lambda^2\times\mathrm{RM},
\end{split}
\end{equation}
where, $\mathrm{e}$ is the charge of an electron; $\mathrm{m_e}$ is the mass of an electron; $\mathrm{c}$ is the speed of light; $\epsilon$ is the permittivity of the medium; $\vec{r}$ is the coordinates of a point in three-dimensional space; $\vec{s}$ is a vector coordinate along the LoS through the plasma, and $\mathrm{RM}$ is called ``Rotation Measure" \citep{Brentjens2005}.
Measuring FR due to the heliospheric and/or CME plasma along the lines of sight to background linearly polarised sources can be a powerful remote sensing tool for measuring these magnetic fields (see \citet{Oberoi2012_FR} for a review). This technique has been successfully used in past; using spacecraft radio beacons \citep[e.g.][etc.]{bird1990, Jensen2013, Wexler2019} and more interestingly also using astronomical sources \citep[][]{Mancuso2000, kooi2017, Kooi2021}. Most of these observations were carried out at higher frequencies and using small FoV instruments, and thus can sample only a small part of the heliosphere at any given time. Moreover, the $\lambda^2$ dependence of FR implies that at high frequencies small change in $\mathrm{RM}$ due to a small change in $\mathrm{\vec{B}}(\vec{r})$ may not produce detectable changes in $\mathrm{FR}$. 

Wide FoV instruments like the MWA can sample large parts of the sky at any given time and can track CMEs as they make their way across the heliosphere. Measurements of FR simultaneously for a large number of pierce points across the CME/heliosphere, open the very exciting possibility of constraining the models for CME/heliospheric magnetic fields using these data \citep{Bowman2013, Nakariakov2015}.  Though scientifically very rewarding and exciting, the following challenges need to be overcome before we can measure the Heliospheric FR using wide FoV instruments:
\begin{enumerate}
    \item The Sun is by far the strongest source in the low radio frequency sky and always makes the dominant contribution to the visibilities measured by large FoV instruments.
    For daytime measurements of the heliospheric magnetic fields, one needs to do high fidelity full Stokes imaging of astronomical sources with polarised flux densities of at most a few $\mathrm{Jy}$ in the presence of the Sun, the flux density of which is four or more orders of magnitude larger.
    \item The instrumental polarisation for wide FoV aperture array instruments is direction-dependent. This instrumental artifact needs to be corrected with high accuracy to get to the heliospheric RM signal \citep[e.g.][]{lenc2017}.
    \item At the meter wavelengths, ionospheric RM is of a similar order of magnitude or larger as compared to the RM due to a CME far away from the Sun (see Fig. 2 of \citet{Oberoi2012_FR}). Precise correction of ionospheric RM variation is, hence, essential for heliospheric FR measurements.
\end{enumerate}

With the availability of P-AIRCARS \citep{Kansabanik2022b, Kansabanik2022c, Kansabanik2022d}, the first of these challenges has now been met (Sec. \ref{sec:P-AIRCARS}). 
High dynamic range imaging provided by P-AIRCARS allows \citet{Kansabanik2022a} to detect numerous background galactic and extra-galactic sources in Stokes I down to flux density of 4.6 $\mathrm{Jy}$ in the presence of the $10^4\ \mathrm{Jy}$ Sun. 
The white dots in Fig. \ref{fig:other_sources} (left panel) are the astronomical sources and the location of the Sun is shown by the red circle. 
The most promising source of linearly polarised emission at these frequencies is however the large angular scale Galactic background \citep{lenc2017}.
We demonstrate that galactic diffuse emission can also be detected in the presence of the Sun (Fig. \ref{fig:other_sources} (right panel)). 
The bright source at the center of the image is the Sun and the extended emission towards the lower left corner is the diffuse emission from the Galactic plane. 
Being able to detect these sources during the daytime marks significant progress toward measuring heliospheric FR during the daytime. P-AIRCARS delivers precise polarisation calibration and produces a high dynamic range of full Stokes images. We are pursuing the objective of demonstrating making similar full Stokes maps of background sources and enabling these heliospheric FR measurements.
With the much-improved sensitivity of the SKAO, it should become possible to routinely perform these observations.

\subsection{Radio Occultation Observations to Probe Solar corona and the Solar Wind}
Radio occultation observations of point sources are an excellent probe of the solar corona and the solar wind. Such observations help in understanding density structure-function, the amplitude of density turbulence, density modulation index, heating rates in the solar wind, dissipation scales among other things. 
When a compact radio source is observed through the foreground solar wind, it can lead to one or more of the following observable consequences : 
(i) the apparent size of the radio source increase due to the scattering of the radio waves in the turbulent medium (referred to as angular broadening or scatter broadening);
(ii) this angular broadening is accompanied by a drop in the apparent peak flux density while conserving the flux density integrated over the entire source;
(iii) the observed anisotropy in angular broadening (typically observed below $10\ R_{\odot}$) carries an imprint of the anisotropies in the coronal and solar wind medium giving rise to the scattering; and
(iv) the position angle (PA; measured from the north through the east) of the measured anisotropies can help constrain the orientation of the magnetic field in the corona. 

Currently, the biggest limitations of this technique come from the limited ability of existing instruments to make sensitive measurements in the proximity of the Sun.
There are few ecliptic sources with flux density large enough to be comfortably imaged as they are occulted by the corona.
The ability to image a large number of background sources in the same field of view as the Sun has already been demonstrated by SKAO precursors like the MWA \citep{Kansabanik2022a}.
Both SKAO-Mid and SKAO-Low promise all of the essential ingredients to extract the numerous scientific benefits from such studies -- higher sensitivity and imaging dynamic ranges and much better angular resolutions.
In the interim significant progress can already be made using observations from precursor facilities like MWA, MeerKAT, and ASKAP.

The favorite source used for such studies thus far has been the Crab Nebula, whose distance of closest approach to the Sun is $\sim 5 R_{\odot}$.
The angular broadening of the Crab Nebula was first reported by \citet{1952Natur.170..319M}. After that many authors have reported such observations at different frequencies (ranging from radio and microwave) and using interferometers with varying baseline lengths \citep{Hew1957,Hew1958,Slee1959,Hew1963,Eri1964,Sastry1974,Ramesh2001, Sas2016, Sas2017, Sas2019, Sas2021}.
The following text showcases the learnings thus far from occultation studies and the expectations from the SKAO.

\subsubsection{Occultation Observations of Crab Nebula}
\begin{figure*}[!ht]
    \centering
    \includegraphics[scale=0.4]{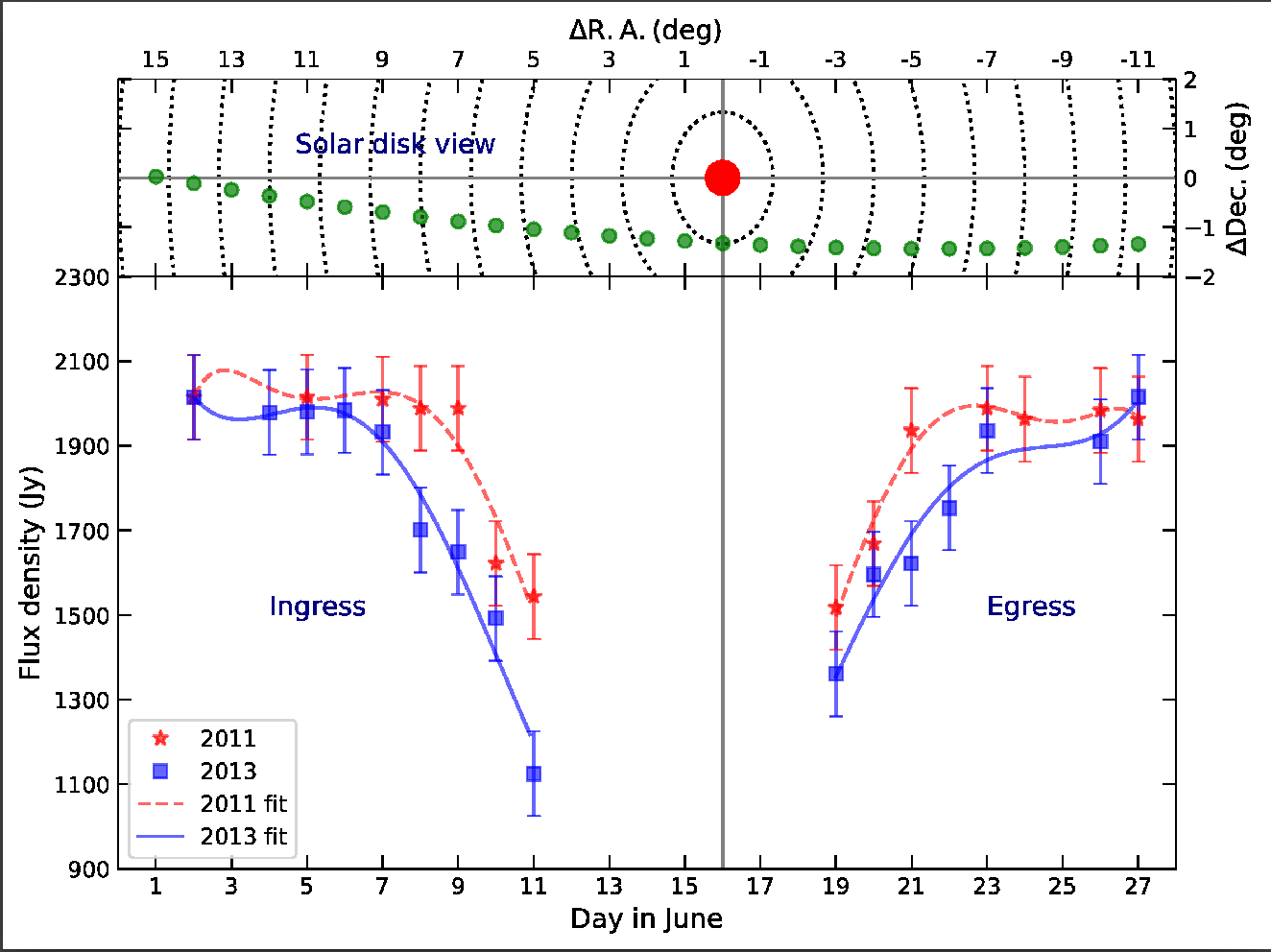}
    \caption{A solar disk view of the Crab Nebula occultation is shown in the upper panel. The red circle indicates the Sun, and the green circles represent the location of the Crab Nebula with respect to the Sun on different days in June. $\Delta$ R.A. and $\Delta $dec. are the offset distances of Crab Nebula from the Sun in right ascension and declination, respectively. The closest dotted concentric circle around the Sun has a radius of $\approx 5~R_{\odot}$. The radii of the circles increase in steps of $5~R_{\odot}$. The lower panel shows the observed flux density of the Crab Nebula on different days during its occultation by the solar corona. The periods before and after June 16th correspond to the ingress and egress, respectively. The symbols `*' and square indicate the observations carried out in 2011 and 2013 June, respectively. The minimum detectable flux density of the Gauribidanur Radioheliograph $\approx 100$ Jy is used as an error bar.}
    \label{fig:crab_sche}
\end{figure*}

The parameters discussed above vary with heliocentric distance and phase of the solar cycle. Such observable quantities provide clues to estimating the various turbulence parameters of the solar corona and solar wind. For example, Figure \ref{fig:crab_sche} shows the observations carried out at GRO in the years 2011 and 2013. 
It is evident that as the Crab Nebula approaches the sun (ingress), the peak flux density decreases until 11 June. 
Crab is not detected from 12--18 June as its flux density drops below the instrument sensitivity threshold.
The flux density increases from 19 June onwards as the Crab Nebula egresses. Note that at its closest, the Crab Nebula is about 5 \rsun from the Sun. 

Using the Very Large Array observations in A configuration (baselines up to 35 km) \citet{Ana1994} observed 
multiple sources with flux densities in the range 0.5--0.9 Jy at 1.3, 2, 3.5, 6, 20, and 90 cm. The solar elongation of these sources ranges between 2--20 $R_{\odot}$. These observations show that angular broadening is evident at microwave frequencies.

\subsubsection{Density Structure Function}
By measuring the peak flux density (V(s)) and zero-baseline flux density (V(0)) or the radio source is observed far from the Sun $\gtrsim 50 R_{\odot}$) one can measure the mutual coherence function ($\Gamma(s)$) and then the density structure function $D_{\phi}(s)$ using \citep{Pro1975} 
\begin{equation}
    D_{\phi}(s)=-2ln[\Gamma(s)]=-2ln[V(s)/V(0)]
\end{equation}

\subsubsection{Amplitude of Turbulence}
By knowing $D_{\phi}(s)$, it is possible to measure the amplitude of the turbulence ($C_N^2$) using the General Structure Function (GSF) \citep{Ingale2015} defined as, 

{\begin{eqnarray}
\label{eq:gsf}
\nonumber
{D_\phi(s)} = \frac{8 \pi^2 r_e^2 \lambda^2 \Delta L}{\rho~ 2^{\alpha-2}(\alpha-2)} {\Gamma \bigg( 1 - {\frac{\alpha-2}{2}} \bigg)}
	    {\frac{C_N^2 (R) l_i^{\alpha-2}(R)}{(1 - f_p^2 (R) / f^2)}} \\
	     {\times \bigg\{ { _1F_1} {\bigg[ - {\frac{\alpha-2}{2}},~1,~ - \bigg( {\frac{s}{l_i(R)}} \bigg)^2 \bigg]} -1 \bigg\}} \, \, \, \, {\rm rad}^{2},
\end{eqnarray}}

where ${ _1F_1}$ is the confluent hypergeometric function, $\lambda$ is the observing wavelength, $R$ is the heliocentric distance (in $R_{\odot}$), $r_e$ is the classical electron radius,  $\Delta L$ is the thickness of the scattering medium, $\rho$ is the anisotropy of the source, $s$ is the baseline length of the interferometer, $\alpha$ is the slope of the turbulent spectrum ($\approx 3$ or $\approx 11/3$), $f_p$ and f are the plasma and observing frequencies, respectively and $l_i$ is the inner scale. 

\subsubsection{Density Modulation Index}
By knowing the $C_N^2$, it is possible to measure the density fluctuations $\delta N_{k_i}$ at the inner scales using \citep{Cha2009}

\begin{equation}\label{eq:deltn}
{\delta}N_{k_i}^2(R) = 4 \pi C_{N}^{2}(R) k_i^{3 - \alpha} e^{-1} \,,
\end{equation}

where $k_i = 2 \pi / l_i$. 

Further, by knowing the ${\delta}N_{k_i}$ and the background electron density 
($N_{e}$), the density modulation index ($\epsilon_{N_e}$) can be measured using, 
\begin{equation}\label{eq:df}
\epsilon_{N_e}(R) \equiv {\frac{\delta N_{k_{i}}(R)} {N_{e}(R)}}. 
\end{equation}

\subsubsection{Proton Heating Rates}
By making use of $\epsilon_{N_e}$, it is possible to measure the proton heating rates in the solar wind. \citet{Cha2009} suggested that density fluctuations at small scales are manifestations of low frequency, oblique ($k_{\perp} \gg k_{\parallel}$), $\rm Alfv\acute{e}n$ wave turbulence and are often referred to kinetic $\rm Alfv\acute{e}n$ waves. 
Note that the quantities $k_{\perp}$ and $k_{\parallel}$ are the components of the wave vector k in perpendicular and parallel directions to the background large-scale magnetic field, respectively.

\citet{Cha2009, Sas2017, Sas2021} envisage a situation where the ``balanced'' counter-propagating $\rm Alfv\acute{e}n$ waves cascade and resonantly damp on the protons at the inner scale and thus heat the solar wind. The proton heating rate (i.e. the turbulent energy cascade rate) at inner scales is, 

\begin{equation}\label{eq:hr}
\epsilon_{k_i}(R)=c_0 \rho_p k_i(R) \delta v_{k_i}^3(R) ~ \rm erg ~cm^{-3}~s^{-1} \, ,
\end{equation}

where, $\rho_p=m_pN_e(R)~\rm g~ cm^{-3}$ with $m_p$ is the proton mass,  $k_i=2 \pi/l_i$ and $\delta v_{k_i}$ are the wavenumber and  magnitude of turbulent velocity fluctuations at inner scales. The dimensional less quantity $c_0$ is assumed to be 0.25.

By knowing the $\epsilon_{N_e}$, one can derive $\delta v_{k_i}$ using the 
kinetic $\rm Alfv\acute{e}n$ wave dispersion relation 
\begin{eqnarray}\label{eq:rmsv}
 \delta v_{k_i}(R)=\Bigg(\frac{1+\gamma_i k_i^2(R) \rho_i^2(R)}{k_i(R) l_i(R)} \Bigg)& \epsilon_{N_e} (R, k_i) v_A(R) \, 
\end{eqnarray}
where the adiabatic index $\gamma_i$ is taken to be 1 and $v_A$ is the $\rm Alfv\acute{e}n$ speed in the solar wind.

\subsubsection{Observations Through CMEs and Coronal Streamers}
\begin{figure*}[!ht]
    \centering
    \begin{minipage}{0.45\textwidth}
            \includegraphics[scale=0.24]{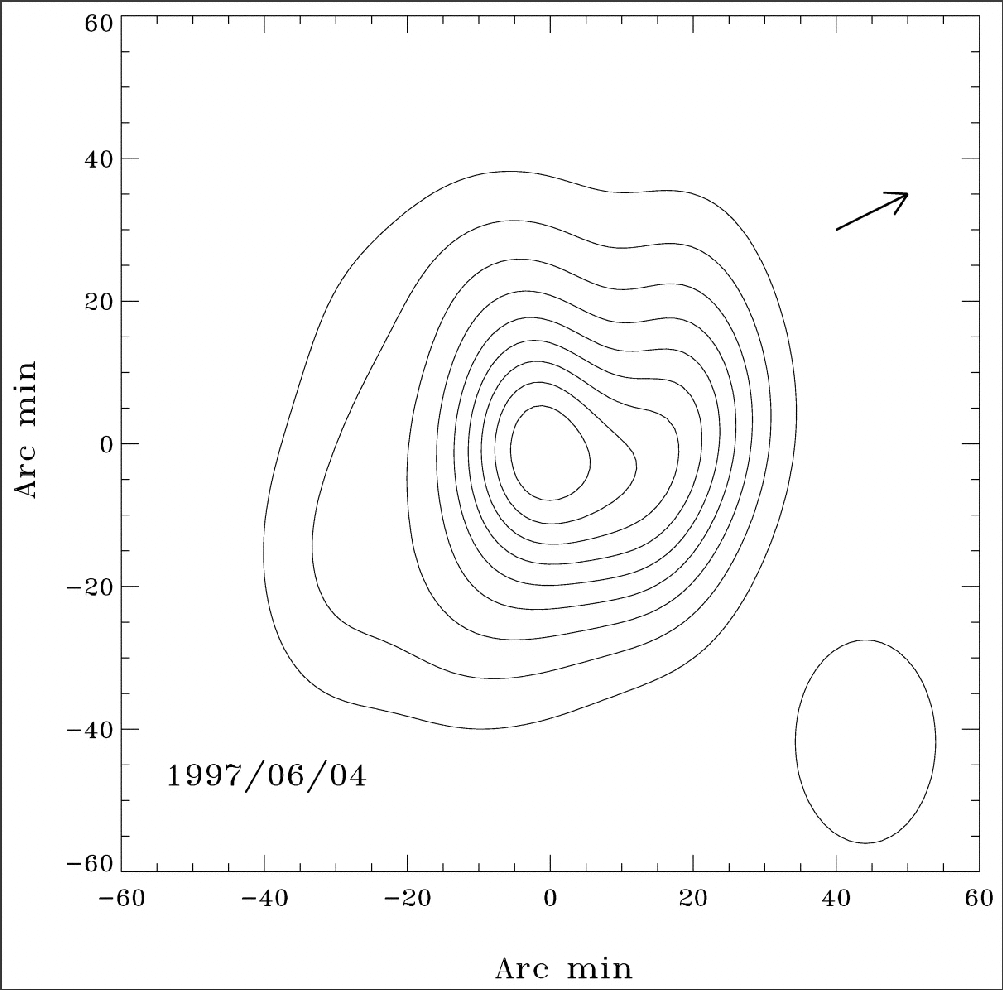}
    \end{minipage}\hfill
    \begin{minipage}{0.45\textwidth}
    \includegraphics[scale=0.23]{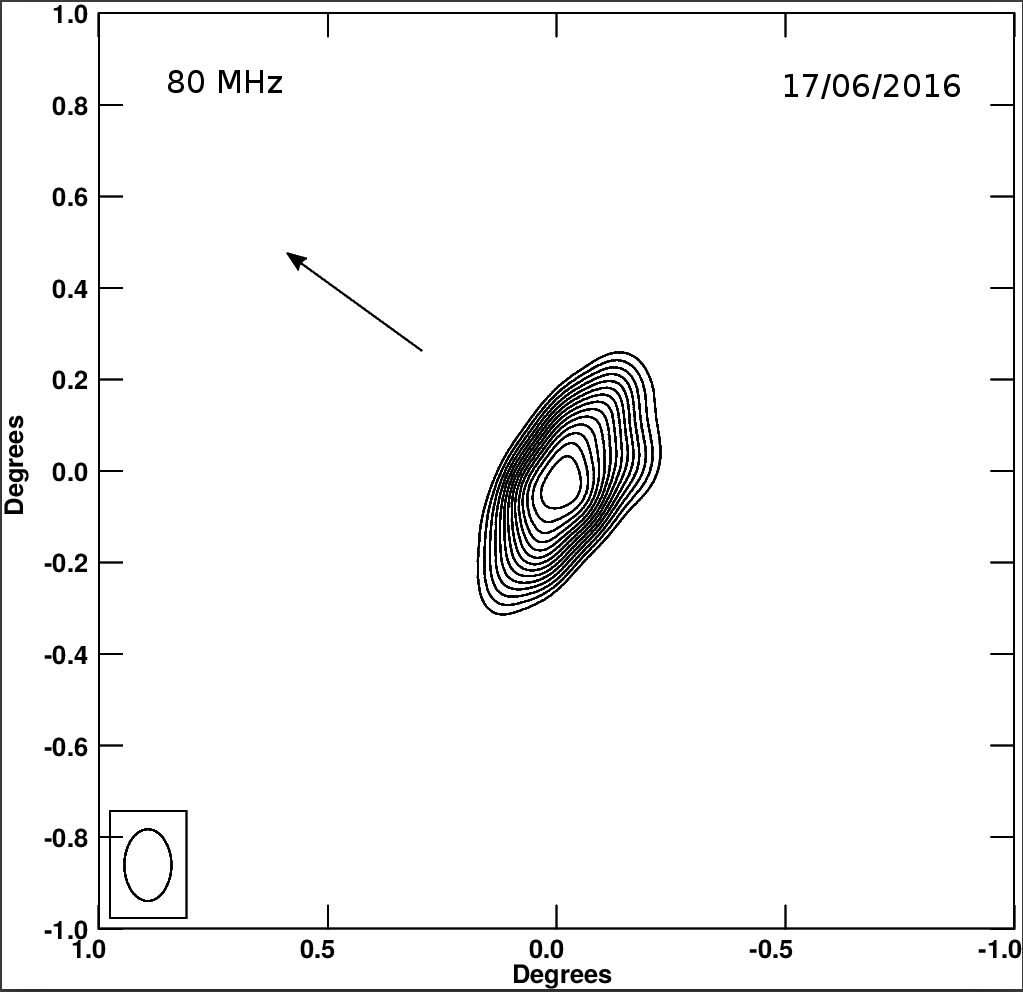}

    \end{minipage}
\caption{Imaging observations of the Crab Nebula by Gauribidanur Radioheliograph. The left panel shows the observations through a Coronal Mass Ejection on 04 June 1997 \citep{Ramesh2001}. The right panel shows the observations through a streamer on 17 June 2016 \citep{Sas2017}. The arrows point in the sun-ward direction.}
\label{fig:crab_streamer}
\end{figure*}

There were attempts to observe the Crab Nebula through CMEs and streamers. 
The Crab Nebula, when observed through a CME at a distance of 41 \rsun on 04 June 1997 \citep{Ramesh2001} is shown a left panel of Fig. \ref{fig:crab_streamer}. 
The right panel of the same Fig. shows the Crab Nebula, as observed through a streamer on 16 June 2016 (10.2 \rsun) \citep{Sas2017}. 
The Figure clearly shows that isotropic angular broadening is observed when the Crab nebula is observed through a CME. In contrast, a clear anisotropic angular broadening is observed in the case of streamers.

\subsubsection{Dissipation Scales}
\begin{figure*}[!ht]
    \centering
    \begin{minipage}{0.45\textwidth}
            \includegraphics[scale=0.855]{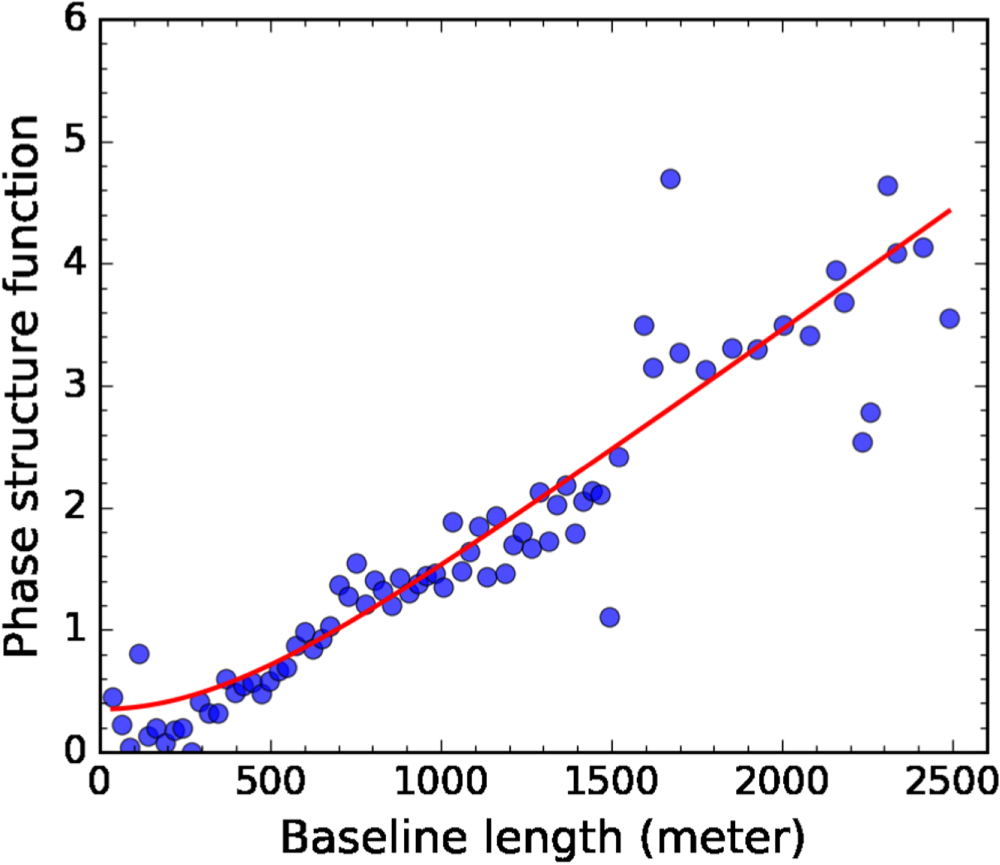}

    \end{minipage}\hfill
    \begin{minipage}{0.45\textwidth}
    \includegraphics[scale=0.90]{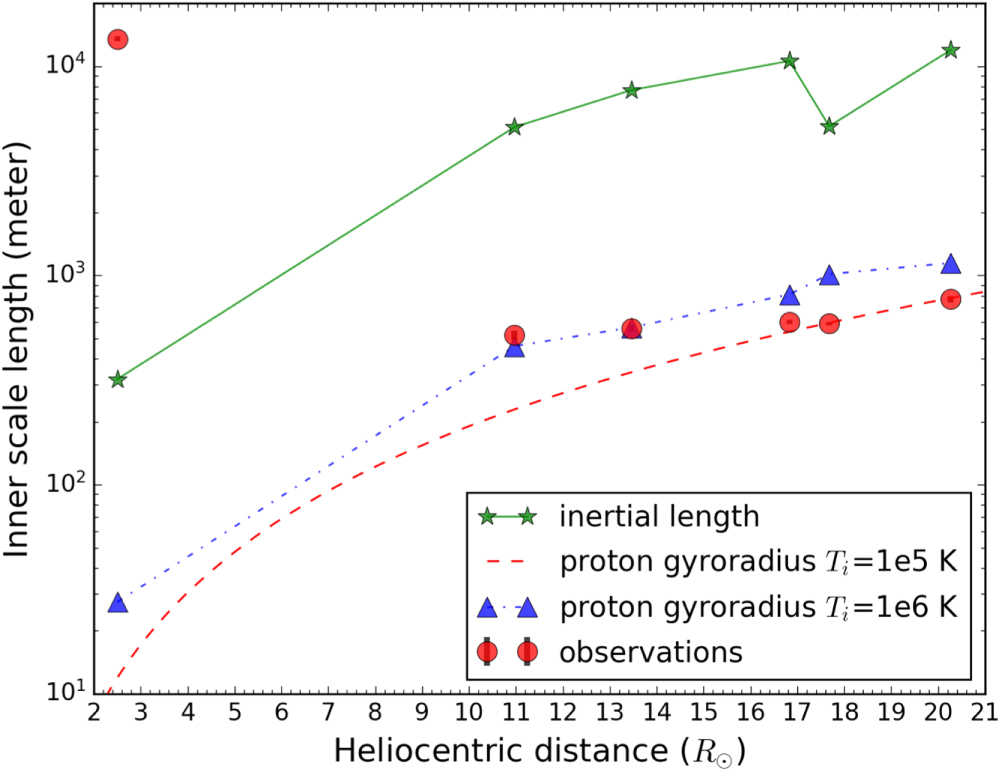}

    \end{minipage}
\caption{In the left panel, blue circles show the structure-function measured from the visibilities vs. the baseline length of the interferometer, and the red curve indicates the GSF model shown in equation \ref{eq:gsf}. In the right panel, red circles show the observationally derived dissipation scales. The solid line indicates the prediction of the proton inertial length model. The dashed line indicates the proton gyroradius computed using a temperature of $10^{5}$ K and the Parker spiral magnetic field model and the dash-dotted line indicates the proton gyroradius using a temperature of $10^{6}$ K and the magnetic field model \citet{Ven2018}.}
\label{fig:dissi}
\end{figure*}

Using multi-baseline observations of radio point sources through the solar corona or solar wind can measure the dissipation scales. Recently \citet{Sas2019} reported the dissipation scales using radio occultation observations from the Gauribindanur Radioheliograph (GRAPH) and the VLA. The calibrated visibilities are averaged over the observation duration and plotted as a function of baseline length (Fig. \ref{fig:dissi}, left panel). These visibilities are fitted with a GSF model shown in Eq. \ref{eq:gsf} for different $l_i$ values. 
The value of $l_i$ which shows the least deviation between the observed and model values is regarded as the dissipation scale for that heliocentric distance.
In the right panel of Figure \ref{fig:dissi}, the red circles indicate the observations, and solid, dashed, and dot-dashed lines are the models for dissipation scales. This study suggests that short baseline observations are crucial in measuring the dissipation scales of the solar corona and solar wind. 

\subsubsection{Variation of the Turbulent Plasma Parameters with Heliocentric Distance and Solar cycle}

The parameters discussed so far in this section, like $C_N^2$, density modulation index ($\epsilon_{N_{e}}$), proton heating rate ($\epsilon_{k_i}$), dissipation scales, etc., all vary with heliocentric distance and the phase of the solar cycle \citep{Sas2016, Sas2019, Sas2021}. A sensitive radio facility like the SKAO, will be able to regularly measure these parameters and establish their dependence on the heliocentric distance and phase of the solar cycle.   

\subsubsection{Magnetic Topology of Solar Wind}
In Figure \ref{fig:crab_streamer}, the arrow is pointing towards the Sun. The right panel shows that the coronal streamer was in the direction opposite the arrow. Also, we know that beyond 10 \rsun, the direction of the streamer will be radially outward direction but the major axis of the Crab Nebula was perpendicular to the direction of the magnetic field direction.
Similar behaviour was also reported by \citet{Ana1994}. 
This has the remarkable implication that the knowledge of the position angle can help ascertain the magnetic topology up to 50 \rsun. Since the SKAO will provide a much higher sensitivity, resolution and imaging dynamic range than the current instrumentation, we can expect to observe many such sources and obtain valuable information on the magnetic topology of the solar wind. 

\subsubsection{Role of the SKAO on Occultation Observation}
As just mentioned, the SKAO is expected to provide a significant improvement over the capabilities of the existing instrumentation and the SKAO precursors and pathfinders. 
By providing the sensitivity and imaging quality to enable observations of a large number of ecliptic sources in the solar vicinity at any given time, these studies will add yet another set of observables to heliospheric studies and space weather. 
These will complement the information obtained from the observations of IPS (Sec. \ref{sec:IPS}) and heliospheric FR (Sec. \ref{sec:helio-FR}), under quiescent conditions and also very interestingly through structures like streamers, interaction regions, and CMEs.
These observations will provide an unprecedented opportunity to characterize the turbulent coronal and solar wind plasma and magnetic fields and potentially 
play a crucial role in the understanding of solar wind heating and acceleration. 
This solar wind is a rich test bed for understanding the properties of magnetohydrodynamic (MHD) turbulence. Using radio occultation observations, we can study and investigate the following aspects.
\begin{itemize}
    \item Understanding the compressibility of solar wind turbulence.
    \item Quantifying the role of radio wave scattering that leads to depressed quiet Sun brightness temperatures at low radio frequencies.
    \item Investigating the impact of solar wind turbulence on the origin, coronal propagation, and the apparent sizes of solar radio burst sources. 
    Being able to disentangle the propagation effects from intrinsic ones will, in-turn, help in building a more detailed understanding of the emission mechanisms involved.
    \item Understanding the range of inner scales/dissipation scales present in the solar wind.
    \item Understanding the role of turbulence in the propagation of Solar Energetic Particles (SEP) through the heliosphere \citep[e.g.][]{Reid2010}.
    \item Carrying out a detailed comparison the various turbulence parameters measured using independent remote sensing methods like occultation studies and in-situ measurements by future instruments similar to Parker Solar Probe, Solar Orbiter, etc.
\end{itemize}

\section{Conclusion}
Indian researchers have been systematically preparing for solar and heliospheric science which will be enabled by the upcoming SKAO.
On the solar physics front, we have been leading the global solar science effort with the only SKAO-Low precursor, the MWA. 
We have successfully met the challenge of building unsupervised automated imaging pipelines for full Stokes snapshot spectroscopic imaging with aperture array instruments like the SKAO-Low. 
The outputs from these pipelines represent the state-of-the-art in metrewave solar imaging, marking improvements of one or more orders of magnitude in dynamic range, polarisation purity, and fidelity. 
Our explorations of the novel phase space made accessible by these images have led to multiple interesting studies.
These science projects have carefully been chosen to match the advantages provided by our high-quality spectroscopic snapshot imaging capability and highlight previously unappreciated or unknown aspects of well-known active emissions or tackle challenging problems like detection of expected but weak gyrosynchrotron emission from CMEs and the discovery of the weakest impulsive nonthermal emissions from the quiet Sun reported yet. 
On the heliospheric physics front, we not only have a rich legacy to build on, but recent developments are also promising to provide the necessary stepping stones to SKAO science.
We are actively engaged in pursuing the science opportunities already made available by the SKAO precursors at SKAO-Low frequencies and plan to increase our engagement with work related to SKAO-Mid frequencies.
Alongside advancing the frontiers of scientific knowledge, we are, on the hand working to develop the human resource with the necessary scientific and technical skills to make the most of the SKAO when it becomes available, and on the other developing state-of-the-art imaging pipelines which will make high quality radio imaging analysis more accessible to the larger solar physics community.


\section*{Acknowledgements}
This scientific work makes use of the Murchison Radio-astronomy Observatory, operated by CSIRO. We acknowledge the Wajarri Yamatji people as the traditional owners of the Observatory site. Support for the operation of the MWA is provided by the Australian Government (NCRIS), under a contract to Curtin University administered by Astronomy Australia Limited. We acknowledge the Pawsey Supercomputing Centre which is supported by the Western Australian and Australian Governments.
DO and DK acknowledges the support of the Department of Atomic Energy, Government of India, under project no. 12-R\&D-TFR-5.02-0700.
This work is supported by the Research Council of  Norway through its  Centres of  Excellence scheme, project number 262622 (“Rosseland Centre for Solar Physics”). AM  acknowledges  support  from  the EMISSA project funded by the Research Council of Norway (project number 286853)
We thank the developers of Python 3 \citep{van1995python,python3} and the various associated packages, especially Matplotlib \citep{Hunter:2007}, Astropy \citep{price2018astropy}, NumPy \citep{Harris2020} and SciPy \citep{Scipy2020}. KSR acknowledges the useful discussions that he had with R. Ramesh, Indian Institute of Astrophysics, Bangalore, India, and Prasad Subramanian, IISER-Pune, India. SM acknowledges partial support by USA NSF grant AGS-1654382 to the New Jersey Institute of Technology.

\vspace{-1em}
\bibliography{references}{}
\bibliographystyle{apj}
\end{document}